\documentclass[sigconf,nonacm]{aamas} 

\usepackage{balance} % for balancing columns on the final page
\usepackage{soul}
\usepackage{url}
\usepackage{amsmath}
\usepackage{booktabs}
\usepackage{algorithm}
\usepackage{algorithmic}
\usepackage{hyperref}
\usepackage{multirow}
\usepackage{dsfont}
\usepackage{subcaption}
\title[Joint Intrinsic Motivation for Coordinated Exploration in MADRL]{Joint Intrinsic Motivation for Coordinated Exploration in Multi-Agent Deep Reinforcement Learning}
\subtitle{Extended version of the extended abstract paper published at AAMAS 2024}

\author{Maxime Toquebiau}
\authornote{Contact author: maxime.toquebiau@gmail.com.}
\affiliation{
  \institution{ECE Paris \& Sorbonne Universit\'{e}, CNRS,\
 ISIR}
  \city{F-75005 Paris}
  \country{France}}
\email{maxime.toquebiau@gmail.com}

\author{Nicolas Bredeche}
\authornote{Authors contributed equally to the paper.}
\affiliation{
  \institution{Sorbonne Universit\'{e}, CNRS, ISIR}
  \city{F-75005 Paris}
  \country{France}}
\email{nicolas.bredeche@sorbonne-universite.fr}

\author{Faïz Benamar}
\affiliation{
  \institution{Sorbonne Universit\'{e}, CNRS, ISIR}
  \city{F-75005 Paris}
  \country{France}}
\email{faiz.ben_amar@sorbonne-universite.fr}

\author{Jae-Yun Jun}
\authornotemark[2]
\affiliation{
  \institution{ECE Paris}
  \city{Paris}
  \country{France}}
\email{jaeyunjk@gmail.com}

\begin{abstract}
Multi-agent deep reinforcement learning (MADRL) problems often encounter the challenge of sparse rewards. This challenge becomes even more pronounced when coordination among agents is necessary. As performance depends not only on one agent's behavior but rather on the joint behavior of multiple agents, finding an adequate solution becomes significantly harder. In this context, a group of agents can benefit from actively exploring different joint strategies in order to determine the most efficient one. In this paper, we propose an approach for rewarding strategies where agents collectively exhibit novel behaviors. We present JIM (Joint Intrinsic Motivation), a multi-agent intrinsic motivation method that follows the centralized learning with decentralized execution paradigm. JIM rewards joint trajectories based on a centralized measure of novelty designed to function in continuous environments. We demonstrate the strengths of this approach both in a synthetic environment designed to reveal shortcomings of state-of-the-art MADRL methods, and in simulated robotic tasks. Results show that joint exploration is crucial for solving tasks where the optimal strategy requires a high level of coordination. 
\end{abstract}


\begin{document}

%%% The following commands remove the headers in your paper. For final 
%%% papers, these will be inserted during the pagination process.

% \pagestyle{fancy}
% \fancyhead{}

\maketitle

\section{Introduction}

One crucial aspect of human intelligence is its ability to act coincidentally with other human beings, to either cooperate or compete in a given task. This has led researchers to study reinforcement learning (RL) in the context of multi-agent systems (MAS), where multiple artificial agents interact with their environment and each other while concurrently learning to perform a task~\cite{Tan1993,Lowe2017:MADDPG}. However, having multiple agents in the environment makes the RL process significantly more difficult for several reasons \cite{Wooldridge2009:IntroMAS}. In particular, the global reward depends on the actions of several independent agents, which makes the search for the optimal joint policy more complicated. 

Recently, multi-agent deep reinforcement learning (MADRL) approaches have combined advancements in RL and deep learning to tackle long-standing problems in MAS such as credit assignment or partial observability~\cite{Lowe2017:MADDPG,Rashid2018:QMIX,Foerster2018:COMA}. These techniques are able to solve very complex multi-agent tasks such as autonomous driving \cite{Shalev2016:AutonomousDriving} or real-time strategy video games \cite{OpenAI2019:DOTA2}. However, major issues still remain with these approaches, such as the problem of relative overgeneralization~\cite{Wiegand2003:RelOvergen,Wei2016:RelOvergen} where agents struggle to find the optimal joint policy because local policies are attracted towards suboptimal areas of the search space. This makes most algorithms inefficient in tasks where the optimal strategy requires strong coordination among agents. Relative overgeneralization can be described as a problem of exploration of the joint-state space: as the success of the MAS depends on the coordination of multiple agents, exploring the joint-observation space is required to discover optimal joint behaviors. In this paper, we address the question of how to explore the joint-state space to efficiently discover superior coordinated strategies for solving the task at hand.

In single-agent RL, the problem of exploration has been studied to solve hard exploration tasks where positive reward signals are very sparse. One solution is to use intrinsic motivation~\cite{Schmidhuber1991,Oudeyer2007:IntrMotiv,lehman2011abandoning} to incite agents to explore unknown parts of the environment. In addition to the environment reward, agents are given an auxiliary reward related to the novelty of encountered states. Maximizing this intrinsic reward leads agents to visit previously unexplored regions of the environment, ultimately discovering new solutions to the task. These methods have shown great success in helping RL agents solve hard exploration tasks \cite{Pathak2017:ICM,Badia2020:NGU}. 

In the multi-agent setting, intrinsic objectives have also been studied to induce different kinds of behaviors in agents such as coordinated exploration \cite{Iqbal2019:MultiExplore}, social influence \cite{Jaques2019,Wang2020:EITI} or alignment with other agents' expectations \cite{Ma2022:ELIGN}. However, previous works have only used local observations to generate intrinsic rewards. With partial observability, local observations often lack crucial information to fully understand the current configuration of the environment. In the context of exploration, an intrinsic reward based only on local observations will lead to each agent exploring their own observation space, without considering the current state of other agents. This can result in inefficient exploration in cooperative tasks where the success of the MAS depends on the coordination of all agents.

In this paper, we introduce a novel multi-agent exploration approach called Joint Intrinsic Motivation (JIM) which can be combined with any MADRL algorithm that follows the centralized training with decentralized execution paradigm (CTDE). JIM exploits centralized information to motivate agents to explore new coordinated behaviors. In order to  compute joint novelty, JIM builds from two state-of-the-art approaches: NovelD \cite{Zhang2021:NovelD} for exploring unknown parts of the environment, and E3B \cite{Henaff2022:E3B} for having more diverse trajectories. Adding this auxiliary reward to the agents' objective incites them to diversify their collective behavior until they have a fair knowledge of the environment and can focus on the main task at hand. 

To demonstrate the advantages of our approach, we first design a simple test environment to showcase a clear example of relative overgeneralization. We show that the state-of-the-art algorithm QMIX \cite{Rashid2018:QMIX} struggles in this scenario and that motivating the exploration of coordinated behavior helps solve the task. Next, we validate these results in a continuous virtual environment, showing that coordination tasks benefit from joint exploration. Finally, further analysis is conducted to confirm the strength and scalability of our approach.

\section{Related Works}\label{sec:RelatedWorks}

In recent years, deep reinforcement learning techniques have been used in the context of MAS to tackle long-standing issues in multi-agent learning. Successful single-agent RL approaches have been adapted to the CTDE framework \cite{Lowe2017:MADDPG,Yu2021:MAPPO}, using a centralized value function to guide the training of decentralized policies. Recent studies have investigated the problem of credit assignment \cite{Foerster2018:COMA} in MADRL, i.e., distributing the global reward among agents based on their participation. Value factorization methods also do this implicitly \cite{Sunehag2018:VDN}, combining the output of local value functions into a centralized one that predicts the current value of the system. In particular, QMIX \cite{Rashid2018:QMIX} uses a separate network to predict the Q-value of the joint action, given the output of local Q-values and the global state of the environment. QMIX has established itself as a long-standing state-of-the-art approach, despite its inherent limitations that several works have tried to surpass \cite{Son2019:Qtran,Rashid2020:WQMIX}. However, MADRL algorithms have been shown to suffer from the problem of relative overgeneralization~\cite{Wiegand2003:RelOvergen,Wei2018:MultiSoftQ}. So far, few works have addressed this problem: Wei et al.~\cite{Wei2018:MultiSoftQ} propose maximum entropy RL to explore the joint-action space, and MAVEN \cite{Mahajan2019:MAVEN} augments QMIX using a hierarchical policy to guide the exploration of joint behaviors. 

A promising approach to overcome relative overgeneralization is to intrinsically motivate agents to explore their environment, ultimately discovering the optimal reward signals. In single-agent RL, curiosity has been defined to help agents solve hard exploration tasks~\cite{Schmidhuber1991,Oudeyer2007:IntrMotiv,lehman2011abandoning} by rewarding the visitation of states considered as novel. For measuring novelty, several methods have used the error of trainable prediction models. The Intrinsic Curiosity Module (ICM) \cite{Pathak2017:ICM} trains a model of environment dynamics and uses the prediction error as a measure of novelty. Random Network Distillation (RND) \cite{Burda2019:RND} uses a target network that produces a random encoding of the state and trains a predictor network to generate the same encoding, the prediction error being the measure of novelty. The idea behind these two approaches is that the prediction models will yield low novelty for states similar to what they have trained on while producing high novelty for unknown parts of the environment. RIDE \cite{Raileanu2020:RIDE} and NovelD \cite{Zhang2021:NovelD} use respectively ICM and RND to compute a reward from the difference of novelty between the next state and the current state, pushing the agents to always seek novel states. Similarly, NGU \cite{Badia2020:NGU} and E3B \cite{Henaff2022:E3B} use clustering techniques to reward states that are distant from previous states. Finally, a similar approach is proposed by AGAC \cite{FletBerliac2021:AGAC} which trains an adversarial policy to predict the main policy’s output, the latter being rewarded with the former’s prediction error. %In our case, we choose to combine NovelD and E3B for their performance and simplicity.

In MADRL, recent works have demonstrated the effectiveness of intrinsic rewards in promoting desirable behaviors in groups of agents. One example is social influence~\cite{Jaques2019,Wang2020:EITI} that rewards agents for performing actions that have a significant impact on other agents. Ma et al.~\cite{Ma2022:ELIGN} propose an intrinsic reward based on the average alignment with other agents' expectations, promoting more predictable behaviors in agents. Lupu et al.~\cite{Lupu2021:TrajeDi} propose to reward policies that perform diverse trajectories in comparison to a population of agents, which is shown to help train agents to be more versatile. Du et al.~\cite{Du2019:LIIR} use intrinsic objectives as a credit assignment technique. Finally, Iqbal and Sha~\cite{Iqbal2019:MultiExplore} propose an approach for coordinated exploration using several metrics for estimating the novelty of observations that depend on all agents' past experiences. However, their model is computationally expensive and does not address the exploration of the joint-observation space, which can be problematic for hard exploration tasks where relative overgeneralization can occur.

In this paper, we address the challenge of relative overgeneralization by rewarding agents for exploring the joint-observation space. In the following sections, we will present the necessary formal background and an overview of the proposed algorithm that implements joint intrinsic motivation.

\section{Background}

\subsection{Dec-POMDP}

To describe cooperative multi-agent tasks, we use the definition of decentralized partially-observable Markov decision process (Dec-POMDP) \cite{Oliehoek2016:DecPOMDP}, defined as a tuple $\langle\mathbf{S},\mathbf{A},T,\mathbf{O},O,R,n,\gamma\rangle$ with $n$ being the number of agents. $\mathbf{S}$ describes the set of global states $s$ of the environment. $\mathbf{O}$ is the set of joint observations, with one joint observation $\mathbf{o}=\{o_1,...,o_n\}\in\mathbf{O}$, and $\mathbf{A}$ the set of joint actions, with one joint action $\mathbf{a}=\{a_1,...,a_n\}\in\mathbf{A}$. $T$ is the transition function defining the probability $P(s'|s,\mathbf{a})$ to transition from state $s$ to next state $s'$ with the joint action $\mathbf{a}$. $O$ is the observation function defining the probability $P(\mathbf{o}|\mathbf{a},s')$ to observe the joint observation $\mathbf{o}$ after taking joint action $\mathbf{a}$ and ending up in $s'$. $R:\mathbf{O}\times\mathbf{A}\rightarrow\mathbb{R}$ is the reward function producing at each time step the reward shared by all agents. Finally, $\gamma\in[0,1)$ is the discount factor controlling the importance of immediate rewards against future gains.

\subsection{Intrinsic rewards}\label{sec:Background:IntrRew}

In Section \ref{sec:RelatedWorks}, we introduced intrinsic motivation as a way to incite agents to actively explore their environment. To this end, at each time step $t$, agents receive an augmented reward $r_t=r^{\text{ext}}_t+\beta r^{\text{int}}_t$, where $r^{\text{ext}}_t$ is the extrinsic reward given by the environment, $r^{\text{int}}_t$ is the intrinsic reward, and $\beta$ is a hyperparameter controlling the weight of the intrinsic reward in the agents' objective.

In this section, we describe three methods of intrinsic rewards from the literature that we will use later in Section~\ref{sec:JIM:IntrRew}. 

\paragraph{Random Network Distillation (RND)} In RND, Burda et al.~\cite{Burda2019:RND} compute novelty using two neural networks with the same architecture: a target network $\phi$ and a predictor network $\phi'$. The target's parameters are initialized randomly and fixed. It takes as input the state $s_t$ and produces a random embedding $\phi(s_t)$. The predictor is trained to output the same embedding, minimizing the Euclidean distance:
\begin{equation}\label{eq:RND}
    RND_t(s_t)=\lVert\phi(s_t)-\phi'(s_t)\rVert_2.
\end{equation}
This distance is used as a measure of the novelty of state $s_t$ and is given as an intrinsic reward to agents.

\paragraph{Novelty Difference (NovelD)} Zhang et al.~\cite{Zhang2021:NovelD} build upon RND to devise a novelty criterion termed NovelD. It is defined as follows:
\begin{multline}\label{eq:NovelD}
    N(s_t, s_{t+1}) = \mathrm{max}[RND(s_{t+1}) - \alpha RND(s_t), 0]\times\\
    \times\mathds{1}\{N_e(s_{t+1})=1\},
\end{multline}
with $\alpha$ a scaling factor and $N_e$ an episodic count of visited states. The first part is the core of the novelty criterion. It uses RND to reward agents for positive gains in novelty between the current and the next states. The second part is an episodic restriction that ensures the reward is given only when state $s_{t+1}$ is observed for the first time in this episode. This restriction limits the use of NovelD to discrete state spaces as it relies on an explicit count of visited states. 

\paragraph{Exploration via Elliptical Episodic Bonuses (E3B)} With E3B, Henaff et al.~\cite{Henaff2022:E3B} propose an episodic bonus based on the position of the observed state with respect to an ellipse that fits all states previously encountered in the current episode. Formally, it is computed as follows:
\begin{equation}\label{eq:E3B}
    b(s_t)=\psi(s_t)^\top C^{-1}_{t-1}\psi(s_t),
\end{equation}
with
\begin{equation}
    C_{t-1}=\sum_{i=1}^{t-1}\psi(s_i)\psi(s_i)^\top+\lambda I,
\end{equation}
where $I$ is the identity matrix and $\lambda$ a scalar coefficient. $\psi$ is an embedding network trained using an inverse dynamics model \cite{Pathak2017:ICM}: embeddings of following states $\psi(s_t)$ and $\psi(s_{t+1})$ are used by a separate neural network trained to predict the action $a_t$ taken between these states. As a result of this training process, $\psi$ encodes parts of the observation that are controllable by the agents (details in~\cite{Henaff2022:E3B}). Intuitively, $b$ can be understood as a generalization of a count-based episodic bonus for a continuous state space. States that are close to previously encountered states in the current episode will yield low bonuses, whereas states that are very different will produce high bonuses.

\section{Algorithm}

In this section, we introduce the Joint Intrinsic Motivation (JIM) exploration criterion for coordinated multi-agent exploration. Firstly, we describe the motivation behind our approach by providing a detailed description of the problem of relative overgeneralization. Then, we define our intrinsic reward and explain how it is used in a multi-agent setting with JIM. %Then, we define the intrinsic reward used for motivating agents to explore a continuous state-space environment in a coordinated fashion. Finally, we explain how this reward is used in a multi-agent setting with JIM. 

\subsection{The challenge of coordinated actions}\label{sec:JIM:Motivation}

\begin{figure}[t]
    \centering
    \subcaptionbox{\label{fig:ro_matrix}}{
        \small
        \begin{tabular}{c|c|c|c|}
            \multicolumn{1}{c}{} & \multicolumn{1}{c}{$A$}  & \multicolumn{1}{c}{$B$}  & \multicolumn{1}{c}{$C$} \\\cline{2-4}
            $A$ & $10$ & $-5$ & $-5$ \\\cline{2-4}
            $B$ & $-5$ & $7$ & $7$ \\\cline{2-4}
            $C$ & $-5$ & $7$ & $7$ \\\cline{2-4}
        \end{tabular}}
    \subcaptionbox{\label{fig:ro_map}}{\includegraphics[width=0.29\textwidth]{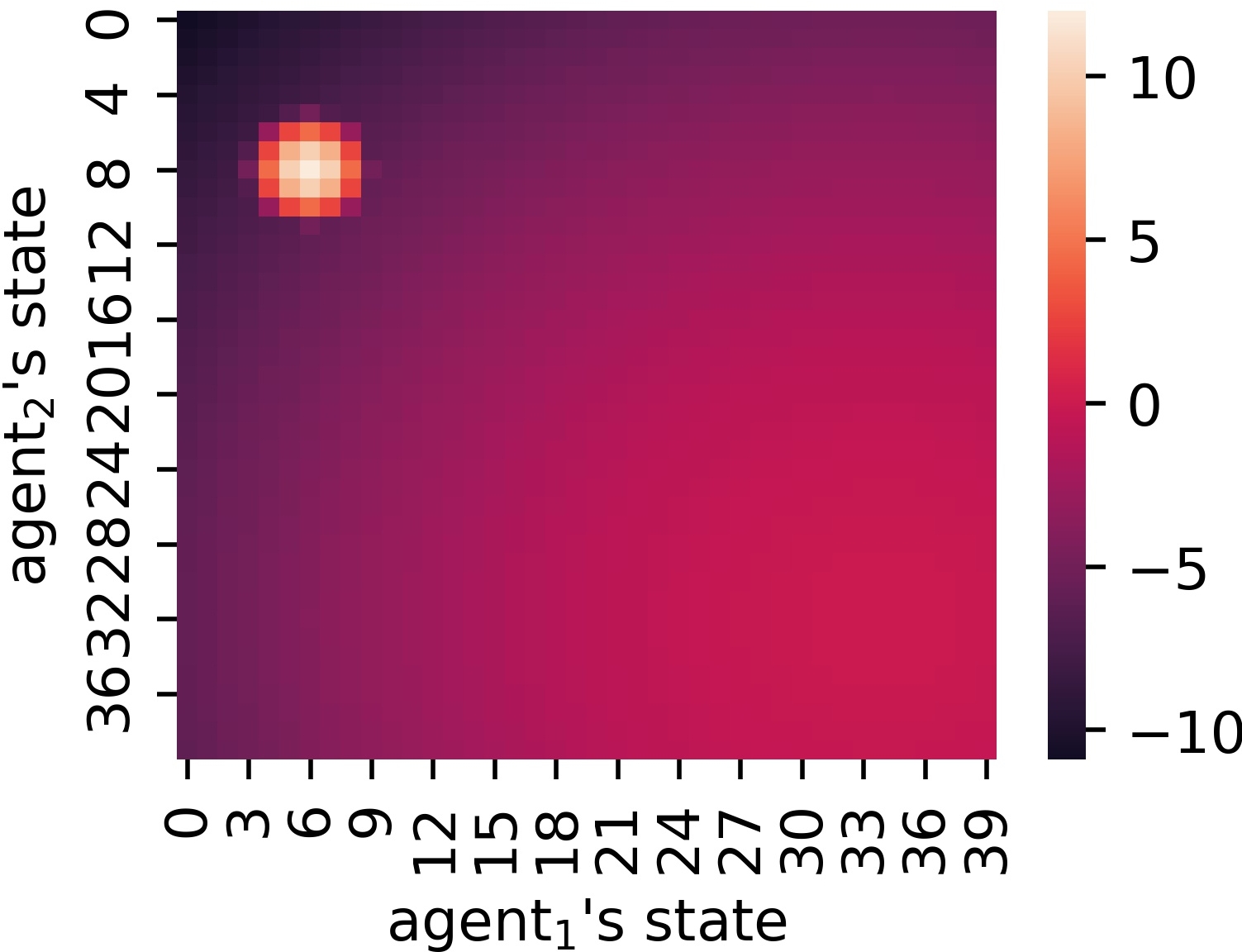}}
    \caption{Two examples of relative overgeneralization: (a) payoff matrix of a social dilemma game, (b) heat-map of the reward function in the $\mathtt{rel\_overgen}$ environment for two agents, with $D=40$ and $\delta=30$. Details in Section~\ref{sec6.1}.}
    \Description{Examples of relative overgeneralization.}
    \label{fig:ro}
    \vspace{-0.3cm}
\end{figure}

Addressing hard exploration environments is challenging because of the sparse positive reward signals that exist to guide the agent's learning process. This becomes even worse with MAS as the completion of a task depends on the actions of multiple independent agents. When strong coordination is needed, agents will struggle to find the optimal strategy and settle for an easier suboptimal joint strategy, which is a problem known as relative overgeneralization~\cite{Wiegand2003:RelOvergen,Wei2016:RelOvergen}. Figure \ref{fig:ro_matrix} provides an example of a social dilemma game where relative overgeneralization occurs. The optimal strategy requires both agents to choose action A. But if only one agent chooses action A, the payoff is very bad. Therefore, agents will independently prefer to take actions B or C, as action A most often leads to sub-optimal outcomes. 

In MAS, this can be seen as a problem of ill-coordinated exploration. As success depends on coordinated behaviors, exploration of joint policies is required in order to discover which ones lead to optimal returns. In the example of Figure \ref{fig:ro_matrix}, exploring independent strategies will lead to ultimately choosing suboptimal actions as they individually may yield better expected returns. On the other hand, we argue that uniformly exploring joint actions would enable agents to choose optimal joint strategies more often and consequently learn more efficient individual behaviors. The approach described in the following two sections implements an algorithm that efficiently rewards agents for exploring the joint-observation space, in order to consistently find optimal strategies. 

\subsection{Double-timescale Intrinsic Reward}\label{sec:JIM:IntrRew}

Similarly to previous works on single-agent intrinsic motivation~\cite{Badia2020:NGU}, we define a novelty metric that combines two exploration criteria working at different timescales:
\begin{itemize}
    \item A \textbf{life-long exploration criterion (LLEC)} that captures how novel is the current observation with respect to all observations since the beginning of training.
    \item An \textbf{episodic exploration criterion (EEC)} that captures the difference between the current observation and all previous observations in the current episode. 
\end{itemize}
Intuitively, the \textit{life-long reward} motivates agents to search for never-experienced parts of the environment. Meanwhile, the \textit{episodic bonus} induces more diverse trajectories. These two elements will feed each other and reinforce agents to efficiently explore their environment. 

Concretely, for each transition from state $s_t$ to the next state $s_{t+1}$, we define the double-timescale intrinsic reward as follows:
% \begin{equation}
%     r_t(s_t, a_t, s_{t+1}) = N_l(s_t,s_{t+1})\times\sqrt{2b(s_{t+1})}
% \end{equation}
\begin{equation}\label{eq:IntrRew}
    r_t(s_t, s_{t+1}) = N_{LLEC}(s_t,s_{t+1})\times N_{EEC}(s_{t+1}),
\end{equation}
with the life-long novelty $N_{LLEC}$ inspired from NovelD~\cite{Zhang2021:NovelD} (see Eq.~\eqref{eq:NovelD}):
\begin{equation}\label{eq:LLEC}
    N_{LLEC}(s_t, s_{t+1}) = \mathrm{max}[RND(s_{t+1}) - \alpha RND(s_t), 0],
\end{equation}
with $\alpha$ a scaling factor and $RND$ the novelty measure (see Eq.~\eqref{eq:RND}). Further, the episodic novelty $N_{EEC}$ uses the bonus from E3B~\cite{Henaff2022:E3B} (see Eq.~\eqref{eq:E3B}):
\begin{equation}\label{eq:EEC}
    N_{EEC}(s_{t+1}) = \sqrt{2b(s_{t+1})}.
\end{equation}

We remove the episodic restriction of NovelD as it relies on an episodic count of visited states. This makes it impractical in a continuous state space, as one state is very unlikely to be visited twice. Instead, we scale the life-long novelty using the elliptical episodic bonus $b$ from E3B~\cite{Henaff2022:E3B}. This bonus acts as an episodic restriction by scaling $N_{LLEC}$ up or down, depending on the novelty of the current state compared to what has been observed during the current episode. As $b$ provides very large bonuses and decreases very fast, we use $\sqrt{2b(s_{t+1})}$ to both smooth out large values and increase small ones. 

Combining these two rewards makes it possible to take the benefits of both. $N_{LLEC}$ pushes agents to explore regions of the state space that are not well-known to agents. Meanwhile, $N_{EEC}$ favors diverse trajectories, inciting agents to always seek new observations during a single episode. As the agents explore their environment, the prediction error of RND (see Eq.~\eqref{eq:RND}) slowly decreases. Thus, $N_{LLEC}$ decreases as well, tending toward zero, allowing agents to progressively focus on the extrinsic reward. Finally, as the episodic restriction does not rely on any explicit count of visited states, it can be used in continuous state spaces.

\subsection{The Joint Intrinsic Motivation algorithm}

\begin{figure}[t]
    \centering
    \includegraphics[width=0.4\textwidth]{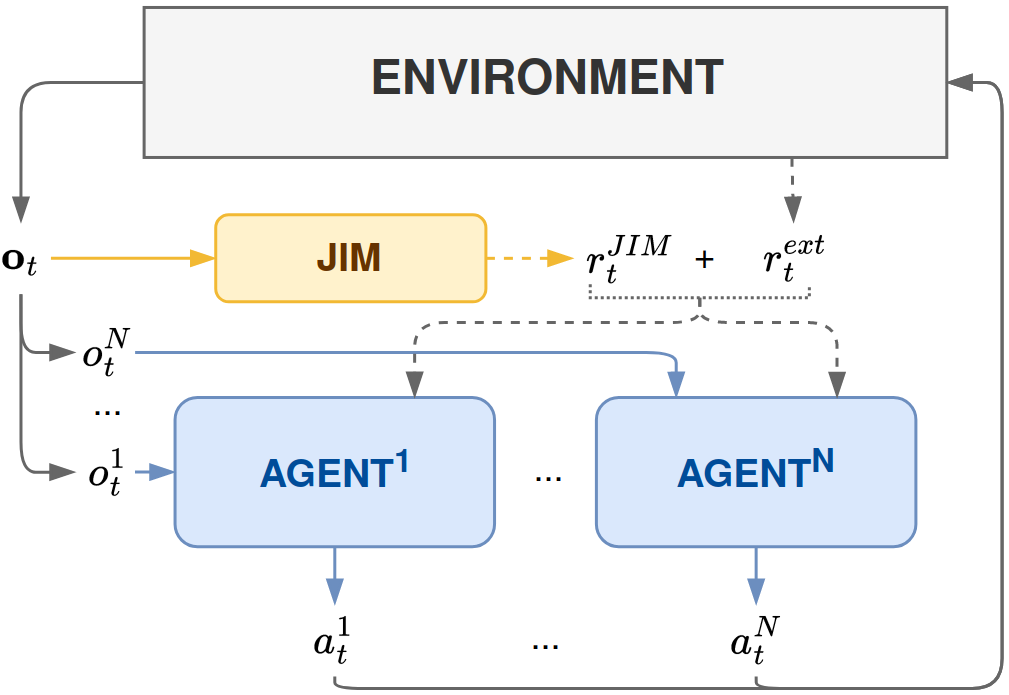}
    \caption{Architecture for the Joint Intrinsic Motivation (JIM) algorithm. JIM has only one intrinsic motivation module for the whole multi-agent system, computing novelty of the joint observation $\mathbf{o}_t$. However, agents only use their local observation to choose their action.}
    \Description[Architecture of Joint Intrinsic Motivation.]{Diagram of the architecture of Joint Intrinsic Motivation. The joint observation is used to generate the intrinsic reward given to all agents.}
    \label{fig:archi}
\end{figure}

Building from the intrinsic reward introduced previously, we propose the Joint Intrinsic Motivation (JIM) algorithm to incite MADRL agents to explore the joint-observation space. At each time step, all agents receive the same global reward $r_t=r^{\text{ext}}_t+\beta r^{JIM}_t$, where $r^{\text{ext}}_t$ is the extrinsic reward given by the environment, $r^{JIM}_t$ is our joint exploration criterion, and $\beta$ is a hyper-parameter controlling the weight of the intrinsic reward. The exploration criterion in JIM uses the double-timescale intrinsic reward defined earlier to compute the novelty of the joint observation:
\begin{equation}\label{eq:JIM}
    r_t^{JIM}(\mathbf{o}_t,\mathbf{o}_{t+1})=N_{LLEC}(\mathbf{o}_t,\mathbf{o}_{t+1})\times N_{EEC}(\mathbf{o}_{t+1}),
\end{equation}
where $\mathbf{o}_t=\{o_t^i\}_{0\leq i\leq N}$, i.e., the concatenation of all local observations. Figure \ref{fig:archi} shows the architecture for JIM. Compared to a local method that would use one intrinsic motivation module per agent, JIM computes only one intrinsic reward. This requires fewer parameters and makes it possible to capture novelty at the team level, rather than at the individual level. As agents are rewarded by the novelty of the joint observation, they will learn to search for new combinations of observations with other agents of the system, rather than only exploring their local-observation space.

As JIM uses joint observations for computing the intrinsic reward, it can be associated with any MADRL algorithm that fits in the CTDE paradigm. These algorithms usually employ a centralized value function \cite{Lowe2017:MADDPG,Rashid2018:QMIX,Yu2021:MAPPO} that looks at the joint observation to predict the value of the agents' actions. Such centralized value functions will be able to associate rewards provided by JIM to new configurations in the joint observation space, thus inducing agents to actively search for these configurations. 

One could note that the joint observation has two notable drawbacks: the number of dimensions grows linearly with the number of agents and there is a risk of capturing redundant information. These issues are both alleviated by using embedding networks to capture the current state of the joint observation into a more condensed latent representation. Both $N_{LLEC}$ and $N_{EEC}$ use embedding networks, respectively $\phi$ and $\psi$ (as described in Section~\ref{sec:Background:IntrRew}), to encode the joint observation. This allows for a more controllable number of parameters in JIM, as only the dimension of the input layers of $\phi$ and $\psi$ depend on the size of the joint observation. Furthermore, embedding networks learn to cast away useless or redundant information in order to produce a faithful, more compact representation of the joint observation. It is important to note that both $\phi$ and $\psi$ were originally used (respectively in RND \cite{Burda2019:RND} and E3B \cite{Henaff2022:E3B}) with raw pixel images as input, showing the significant dimensionality reduction capabilities of these techniques.

%Another option would be to use a global state of the environment comprising all information available in the environment in place of the joint observation. However, not all environments are able to provide such a global state. Thus, we choose not to pursue this idea and stick with the concatenation of all local observations.

\section{Implementation details}

\begin{figure*}
     \centering
     \begin{tabular}{c c c}
        \includegraphics[width=0.3\textwidth]{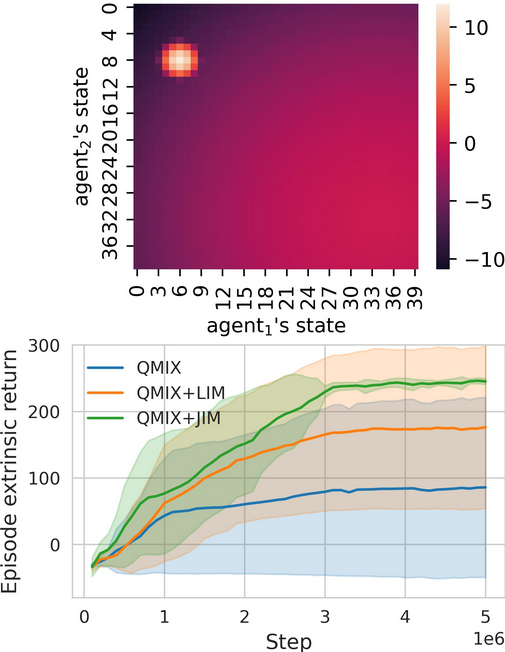} & \includegraphics[width=0.28\textwidth]{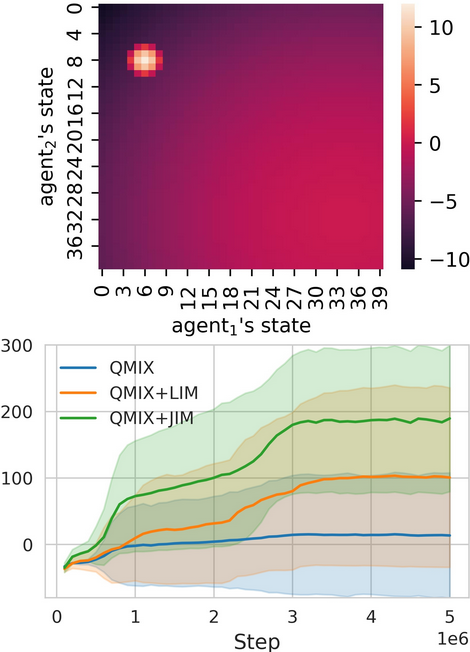} & \includegraphics[width=0.28\textwidth]{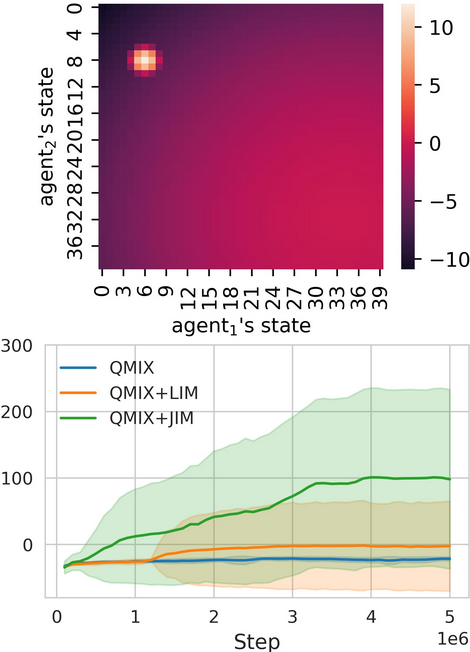} \\
        (a) easy ($\delta=30$) & (b) hard ($\delta=40$) & (c) very hard ($\delta=50$)
     \end{tabular}
     \caption{Performance of variants of QMIX in the $\mathtt{rel\_overgen}$ environment, with three levels of difficulty. On top, we show the heat maps representing the reward function in each instance, where the difficulty is dictated by the width coefficient of the optimal reward spike $\delta$ (as defined in Eq. \eqref{eq:RelOvergen}). Increasing $\delta$ leads to a smaller optimal reward spike. Below is shown the performance during training of QMIX with no intrinsic reward (QMIX), local intrinsic motivation (QMIX+LIM), and joint intrinsic motivation (QMIX+JIM) (mean and standard deviation shown for 15 runs each). We see that a slight decrease in the size of the optimal reward spike results in a considerable increase in the difficulty of the task.}
    \Description[Training curves in the relative overgeneralisation scenario.]{Training curves in the relative overgeneralisation scenario. Three modes of difficulty are studied, each with a different size for the optimal reward spike: the narrower the spike, the harder the task. The curves show that Joint Intrinsic Motivation always improves the performance of QMIX and is better than Local Intrinsic Motivation.}
     \label{fig:ro_results}
\end{figure*}

%As previously said, JIM can be used to augment any MADRL approach that fits in the CTDE paradigm. 
In the next section, we use JIM with QMIX~\cite{Rashid2018:QMIX}. We use the default QMIX architecture and hyperparameters, along with prioritized experience replay~\cite{Schaul2016:PER}. In all experiments, we compare three algorithms: 
\begin{itemize}
    \item \textbf{QMIX+JIM}, augmenting QMIX with joint exploration, as shown in Figure~\ref{fig:archi} and described in Section~\ref{sec:JIM:IntrRew}.
    \item \textbf{QMIX+LIM}, a degraded version of QMIX+JIM where local (rather than global) intrinsic motivation is used. Each agent generates its own intrinsic reward based solely on its local observation, using the same reward definition as JIM (see Section~\ref{sec:JIM:IntrRew}). The architecture for LIM (Local Intrinsic Motivation) is described in Appendix \ref{app:LIM}.
    \item The original state-of-the-art \textbf{QMIX} algorithm~\cite{Rashid2018:QMIX} with no intrinsic motivation, used as a baseline. 

\end{itemize}
Note that the only difference between these three algorithms is the definition of the reward function given to each agent during training. The actual training and execution algorithms are identical.

To ensure a fair comparison between JIM and LIM, we use different values for some specific hyperparameters (e.g., dimension of the hidden layers) in the two versions in order for them to have a similar number of trainable parameters. All hyperparameters used in our experiments are listed in Appendix \ref{app:hpp}\. The code used to run all experiments is freely available online\footnote{https://github.com/MToquebiau/Joint-Intrinsic-Motivation}.

\section{Experiments}

In this section, we present a set of experiments to evaluate the exploration criterion of JIM when used along the state-of-the-art QMIX algorithm \cite{Rashid2018:QMIX}. First, we show the results in a synthetic discrete environment where the problem of relative overgeneralization can be artificially tuned and observe that JIM helps alleviate this issue. Then, we test our approach on pseudo-realistic robotic tasks in a continuous environment and show that exploring the joint-observation space helps solve cooperative tasks. Next, we present an ablation study by comparing JIM with two simpler versions that each lack one of the two exploration criteria described in Section~\ref{sec:JIM:IntrRew}, showing the advantage of combining the two. Finally, we show in Section~\ref{sec:Exp:scaling} that JIM remains relevant when problems are scaled up.

\subsection{Addressing relative overgeneralization}
\label{sec6.1}

\subsubsection{Environment definition}

To demonstrate how joint exploration helps solve the problem of relative overgeneralization, we design a simple test environment that expands the example shown in Figure \ref{fig:ro_matrix}. In this environment called $\mathtt{rel\_overgen}$, two agents can move on a discrete one-dimensional axis with $D$ possible positions. The two agents are denoted by their position, namely $\mathrm{x}$ (for the first agent) and $\mathrm{y}$ (for the second agent). At each time step, agents observe their position as a one-hot vector (e.g., for agent x, $o^\mathrm{x}_t=\{o^{\mathrm{x},i}_t=1\ \text{if}\ \mathrm{x}=i,\ 0\ \text{otherwise}\}_{0\leq i<D}$) and can choose between three actions: move in one direction or the other, or stay in position. They receive a reward corresponding to their combined position:
\begin{align}\label{eq:RelOvergen}
    \begin{split}
        r^{\text{ext}}_t(x,y;\delta)=\mathrm{max}\Bigl(&R^+-\frac{\delta}{D}\bigl[(x-r^+_x)^2+(y-r^+_y)^2\bigr],\\
        &R^--\frac{1}{8D}\bigl[(x-r^-_x)^2+(y-r^-_y)^2\bigr]\Bigr).
    \end{split}
\end{align}
The result of this formula is displayed in Figure \ref{fig:ro_map}. The reward combines two hyperboles in opposite corners: one narrow that culminates at $R^+$ at position $(r^+_x,r^+_y)$, and another much wider that plateaus at $R^-$ at position $(r^-_x,r^-_y)$. We set the optimal reward $R^+$ to $12$ and the suboptimal $R^-$ to $0$. The width of the optimal reward spike is controlled by the parameter $\delta$: a higher $\delta$ value yields a narrower spike. 

The goal of the agents is to find where to go to maximize the global reward. The wide suboptimal hyperbole is deceptive as it is an obvious path for agents to minimize their loss. The optimal reward spike is difficult to find because it covers a small portion of the state space, but it guarantees much greater returns. We can vary the difficulty of the task by changing the width of this optimal reward spike: the narrower the spike, the harder it is to find.

In this environment, we expect MADRL methods to struggle to find the optimal reward spike. Exploring local states could help but would not be sufficient to consistently solve the task. As the dimension $D$ of the local-state space is fairly small, novelty rewards will quickly vanish and will not help agents find the optimal reward spike. Exploring the joint-observation space adequately is required in order to consistently find optimal rewards. As JIM will reward exploration until all combined positions $(x,y)$ are visited several times, agents will visit the optimal reward spike more often, thus helping them to learn the optimal coordinated strategy.
% \vspace{-0.2em}
\subsubsection{Results}

The results shown in Figure \ref{fig:ro_results} confirm the hypotheses formulated in the previous section. We show the performance of QMIX, QMIX+LIM, and QMIX+JIM across 15 independent runs each. Further, we present results in three difficulty levels dictated by the width of the optimal reward spike. The results clearly demonstrate the importance of exploring the joint-state space. QMIX alone manages to get a positive reward on the easy scenario, but its performance is both lower and with a larger standard deviation compared to the two other algorithms. In the harder scenarios, QMIX's performance degrades strongly, never finding any positive reward in the hardest case. JIM clearly improves the performance. In the easy scenario, QMIX+JIM consistently goes for the optimal reward spike. In the harder settings, it still performs well on average, even in the "very hard" scenario where the optimal reward spike covers only 0.013\% of all combined positions. The results of QMIX+LIM show that exploring the local-observation space helps agents find the optimal reward spike more often. However, it performs worse than JIM as it does not ensure that all combined positions are sufficiently explored. This shows that exploring the joint-observation space is crucial to allow agents to discover optimal coordinated behaviors.

\subsection{Coordination tasks in a continuous environment}

\subsubsection{Environment definition and setups}

Next, we study how JIM scales to more realistic continuous environments and more complex tasks. We use the multi-agent particle environment\footnote{https://github.com/openai/multiagent-particle-envs} (MPE) \cite{Mordatch2018,Lowe2017:MADDPG} to simulate cooperative robotic tasks that require a high degree of coordination. The state space of MPE is continuous: in our setups, agents receive as observation a vector with their position in the two-dimensional space and, for all the other entities in the environment, their relative position and velocity. Agents navigate in a closed two-by-two-meter area by choosing between five discrete actions: move in any four cardinal directions or stay in place. 

\begin{figure}[t]
    \centering
    \setlength{\fboxsep}{0pt}
    \setlength{\fboxrule}{1pt}
    \subcaptionbox{\label{fig:env_push}}{\fbox{\includegraphics[width=0.15\textwidth]{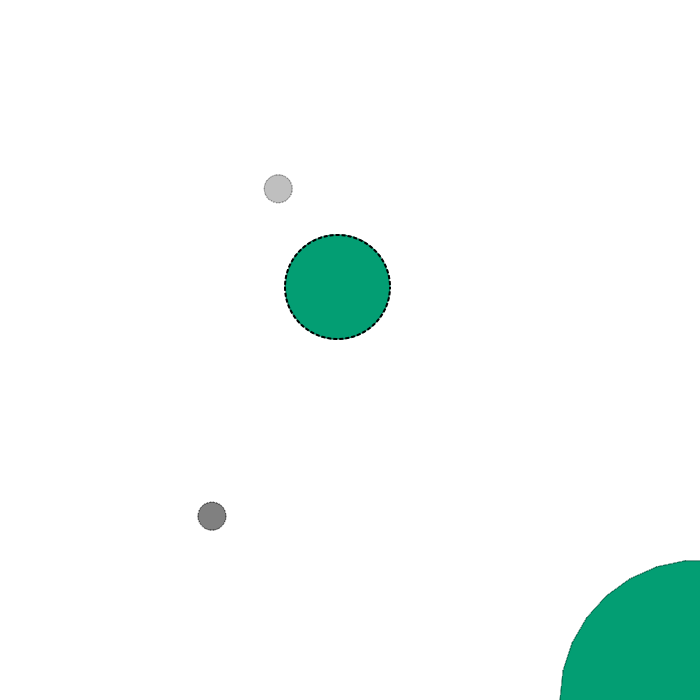}}}
    \subcaptionbox{\label{fig:env_button}}{\fbox{\includegraphics[width=0.15\textwidth]{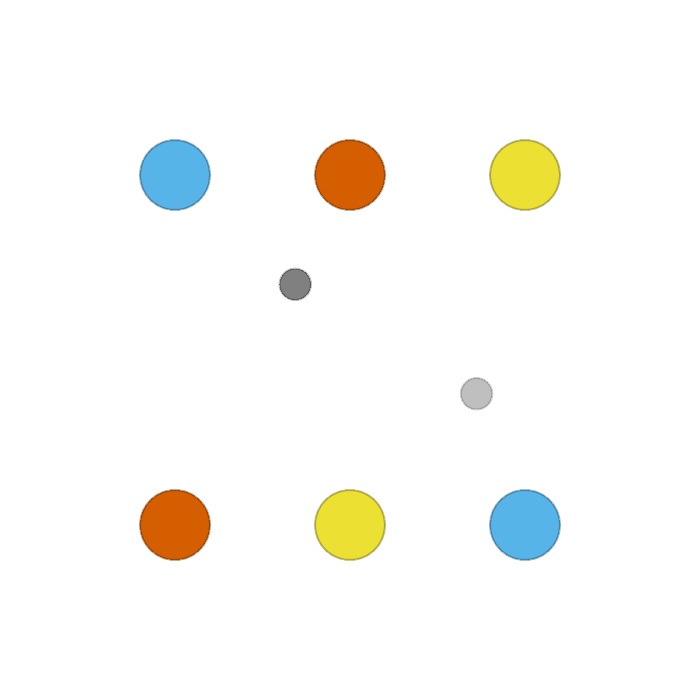}}}
    \caption{Screenshots of our custom tasks in the MPE, agents are the small grey circles. (a) \textit{cooperative box pushing scenario}: agents must deliver the object (green circle in the middle) to the landmark in the bottom right corner. (b) \textit{coordinated placement scenario}: agents must navigate to position themselves on the colored circles representing landmarks.}
    \Description[Tasks in the Multi-agent Particle Environment.]{Screenshots of the two tasks in the two-dimensional Multi-agent Particle Environment. The first one shows the cooperative box pushing task, where two agents navigate around an object that can be pushed toward a landmark. The second one shows the coordinated placement task, with two agents and two sets of three colored landmarks.}
    \label{fig:envs}
\end{figure}

The first task is a cooperative box-pushing task that requires agents to push an object and place it on top of a landmark. Figure \ref{fig:env_push} shows a screenshot of this scenario. At the start of each episode, the landmark is randomly placed in any one of the four corners. The initial positions of the agents and the object are randomly set. If the agents manage to push the object and place it on the landmark, the episode ends and they receive a reward of +100. Agents also get a small penalty of -1 at each time step to reward faster strategies. This reward is purposefully defined to be very sparse, in order to study how exploration helps in this kind of situation. 

\begin{figure}[t]
    \centering
    \includegraphics[width=0.35\textwidth]{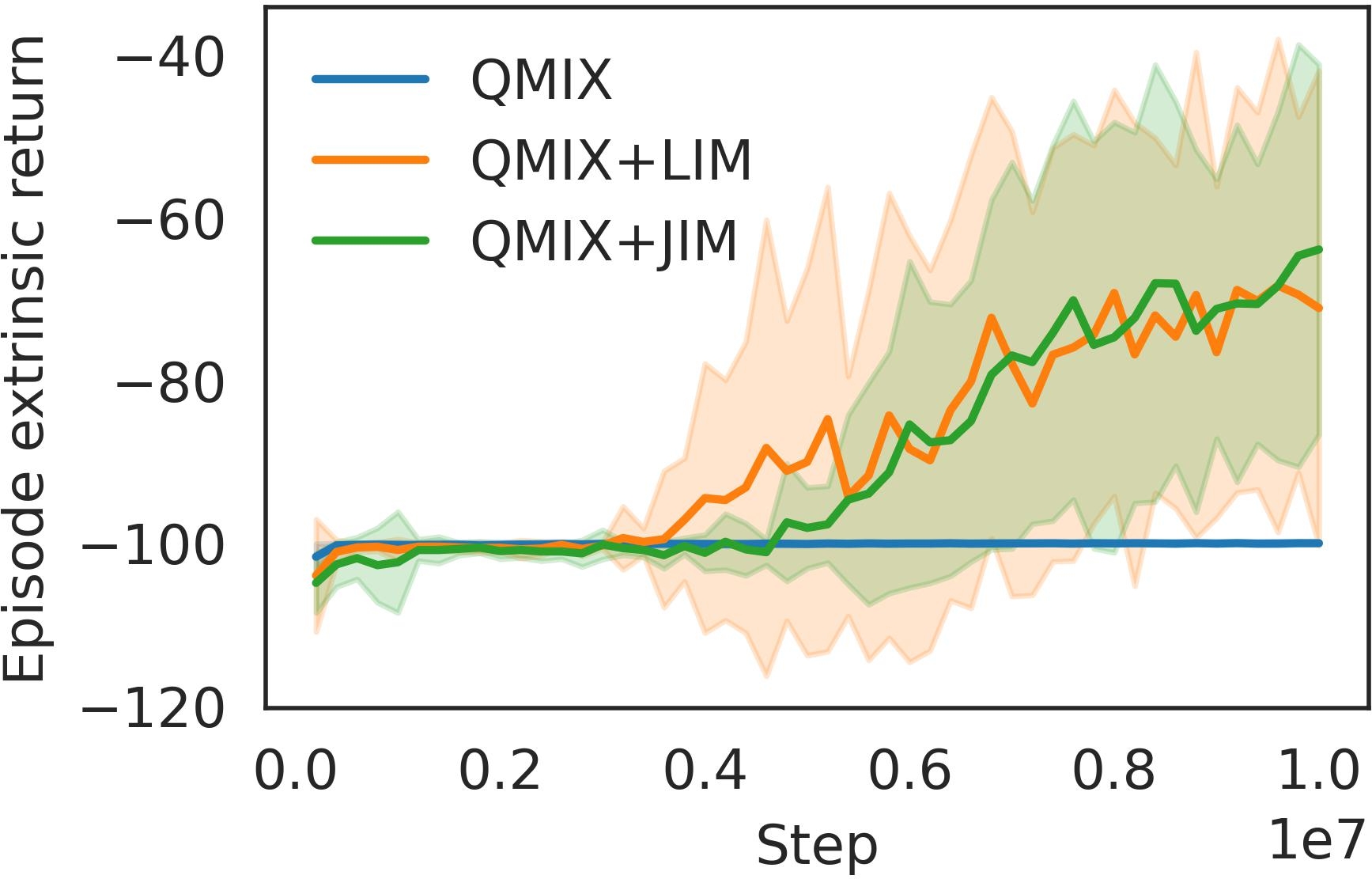}
    \caption{Training curves of the three variants of QMIX in the cooperative box pushing task, with the mean and standard deviation across 11 runs each.}
    \Description[Training curves in the cooperative box pushing task.]{Training curves in the cooperative box pushing task. QMIX alone stays at a reward of -100 during the whole training, while both Joint and Local Intrinsic Motivation slowly converge towards -70.}
    \label{fig:push_results}
\end{figure}

The second scenario is a cooperative placement task where agents must position themselves over landmarks in order to maximize their reward. As shown in Figure \ref{fig:env_button}, there are two sets of three colored landmarks. The reward given at each time step depends on the placement of the agents on the landmarks. The optimal state is having both agents placed on the red landmarks, yielding a reward of +10 at each time step. The blue and yellow landmarks act as deceiving rewards, yielding much smaller rewards (+2 for blue, +1 for yellow). To increase the deceiving aspect of the blue and yellow landmarks, we also reward agents collectively by +0.5 if only one of them stands on one of these two colors. This leads to a need for coordination between agents, as they will locally find that going on blue or yellow landmarks systematically leads to a small reward. Only if agents explore their environment well, will they discover that they need to be both on red to get the optimal reward signal. Importantly, this scenario features partial observability, with agents only having information about entities close to them (less than sixty centimeters from them), the others being masked with neutral values. This means that agents do not necessarily see which landmark the other agent goes to\footnote{See Appendix \ref{app:mpe} for a detailed description of these two tasks.}. 

\subsubsection{Results}

\begin{figure}[t]
    \centering
    \subcaptionbox{\label{fig:button_curves}}{\includegraphics[width=0.295\textwidth]{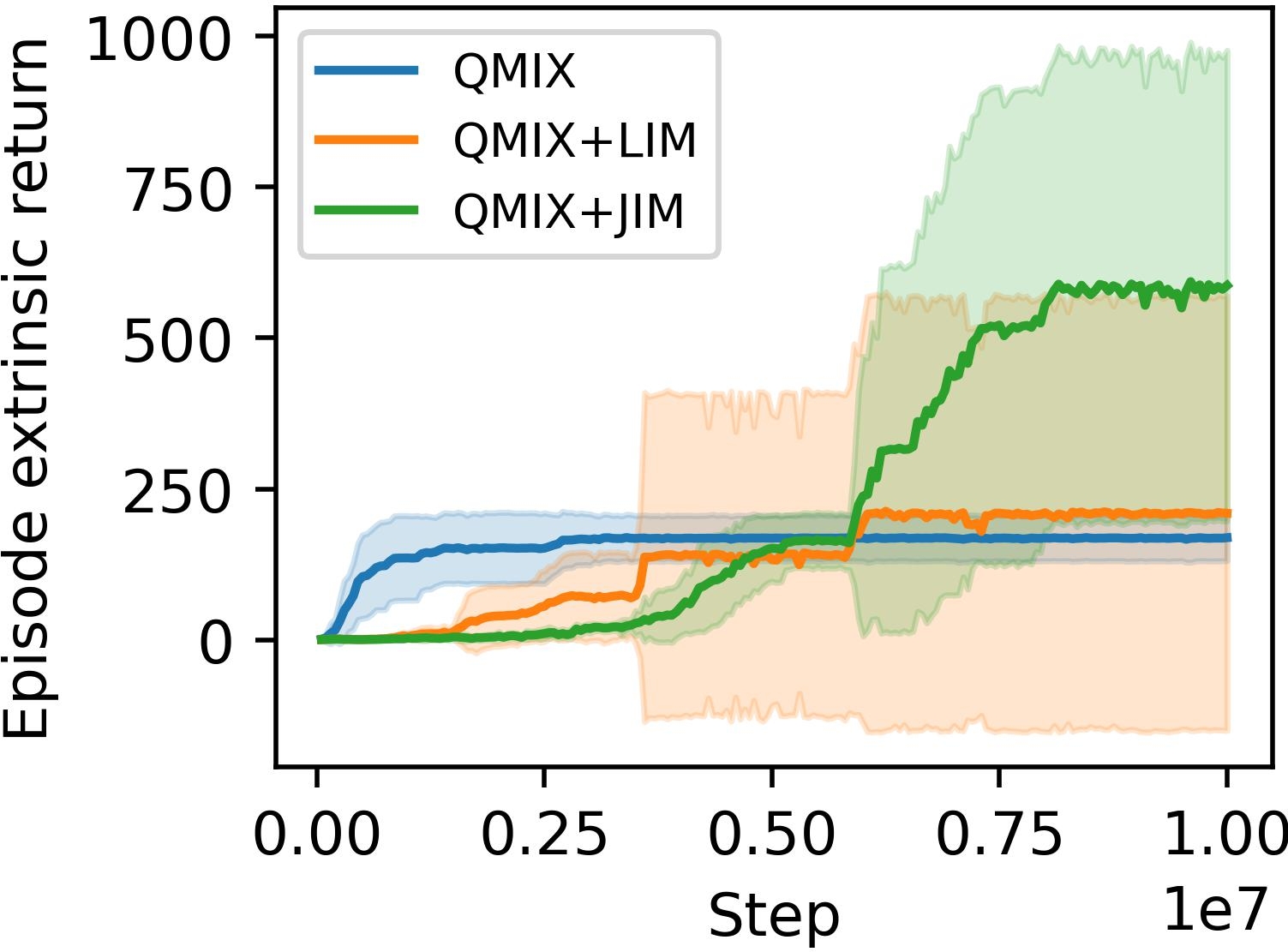}}
    \subcaptionbox{\label{fig:button_last}}{\includegraphics[width=0.175\textwidth]{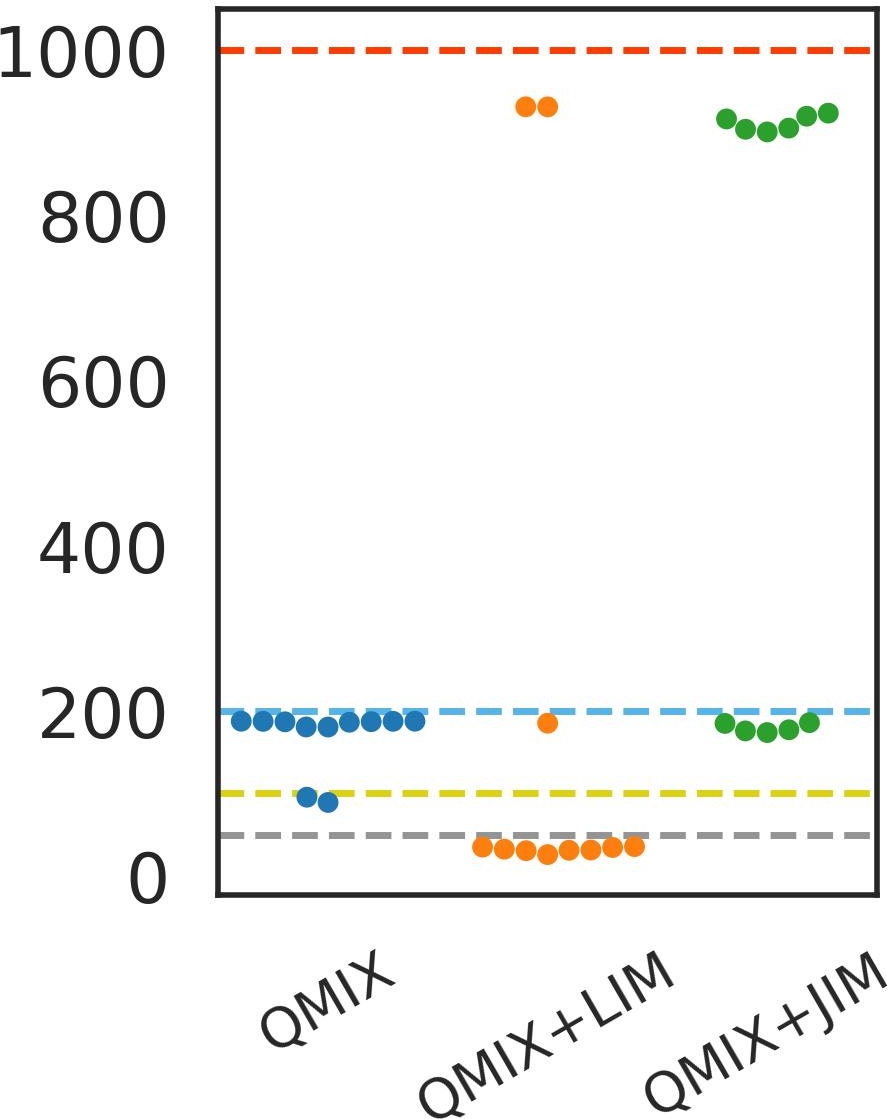}}
    \caption{Performance of the three variants of QMIX in the coordinated placement task. (a) shows the training curves with the mean and standard deviation across 11 independent runs each, while (b) displays the performance of each independent run at the last iteration of training. Dashed lines on (b) indicate the optimal (unattainable) level of return obtained with different strategies: red, blue, and yellow lines represent the return obtained if both agents are on landmarks of the related color during 100 steps (the duration of an episode), and the grey line similarly describes only one agent being either on blue or yellow.}
    \Description[Training curves in the coordinated placement task.]{Two graphs showing training curves in the coordinated placement task and the final reward level obtained by each independent runs (there are 11 for each trained version). Joint Intrinsic Motivation is clearly better, converging the optimal strategy more than half of the time.}
    \label{fig:button_results}
\end{figure}

Results of training in the cooperative box pushing scenario are shown in Figure \ref{fig:push_results} (median and confidence interval shown for 11 runs each). First, we observe that QMIX alone performs very poorly as it is unable to find the solution to the task. The high sparsity of the reward function makes it impossible for agents to discover the objective with random exploration of the environment. Second, we see that JIM and LIM achieve similar levels of performance. While coordination can help agents perform well, it is actually not a requirement for this task. In fact, one agent alone is able to push the object and place it on the landmark. Thus, exploring the space of joint configurations is not helpful in this scenario. This shows however that actively exploring the environment is crucial in tasks where the reward function is very sparse. 

\begin{figure}[t]
    \centering
    \setlength{\tabcolsep}{1pt}
    \begin{tabular}{cccc}
        \includegraphics[width=0.11\textwidth]{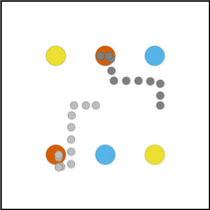} & \includegraphics[width=0.11\textwidth]{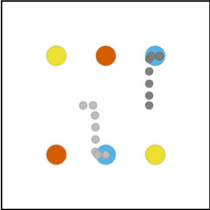} & \includegraphics[width=0.11\textwidth]{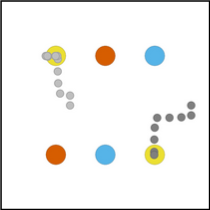} & \includegraphics[width=0.11\textwidth]{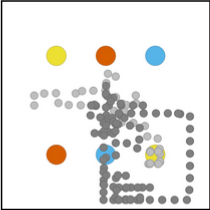} \\
        (a) & (b) & (c) & (d)
    \end{tabular}
    \caption{Examples of trajectories in the coordinated placement task. They present different levels of strategy, with (a) > (b) > (c) > (d): (a) optimal strategy with both agents on red, (b) both on blue, (c) both on yellow, and (d) one on blue/yellow.}
    \Description[Examples of trajectories in the coordinated placement tasks.]{Examples of trajectories in the coordinated placement tasks. There are four examples, showing four levels of strategies, from best (agents go both on red landmarks) to worst (only one agent is on a yellow landmark, the other navigates around the landmark).}
    \label{fig:traj}
\end{figure}

Conversely, experiments in the coordinated placement task de-\\monstrate well the importance of exploring jointly. Figure~\ref{fig:button_curves} shows the training curves of QMIX, QMIX+LIM, and QMIX+JIM, with 11 independent runs each. Figure \ref{fig:button_last} shows the performance of each run at the last iteration of training. The colored dashed lines give an insight into the level of strategy learned by each run. These levels of strategy can be visualized with example trajectories displayed in Figure \ref{fig:traj}. QMIX alone almost always goes for the blue landmarks, while sometimes settling for the yellow ones. This indicates that without actively exploring the environment, QMIX gets stuck because of deceptive rewards and is unable to find the optimal strategy. While QMIX+LIM seems slightly better than QMIX on the training curves, the individual run performance shown in Figure \ref{fig:button_last} shows that LIM arguably performs worse. Two runs manage to find the optimal strategy, but LIM often performs poorly with only one agent on a blue or yellow landmark. This demonstrates that exploring the space of local observations can be helpful, as it pushes agents to explore the environment. However, exploring local observations can also be misleading as they do not contain all the information about the current state of the environment. With JIM, exploring the joint-observation space clearly improves the quality of the chosen strategies. More than half of the time, QMIX+JIM finds the optimal reward signal and learns an effective strategy to go on red landmarks, showing that JIM allows for more efficient exploration of coordinated behaviors. When agents do not find the optimal strategy, they stick with the best sub-optimal strategy to go both on blue. This shows that agents benefit from exploring the space of joint observations as they are directly linked to the obtained reward, whereas local observations lack crucial information to understand the global reward. 

Another advantage of JIM is the simplicity of its architecture. While LIM and similar approaches in previous works \cite{Iqbal2019:MultiExplore,Du2019:LIIR,Wang2020:EITI} require computing one intrinsic reward for each agent, JIM only computes one intrinsic reward for the whole group of agents. This makes JIM significantly more efficient to run, with LIM being approximately 24\% slower than JIM to train (see Appendix \ref{app:train_time} for details). 

\begin{figure}[t]
    \centering
    \subcaptionbox{\label{fig:ablation_curves}}{\includegraphics[width=0.308\textwidth]{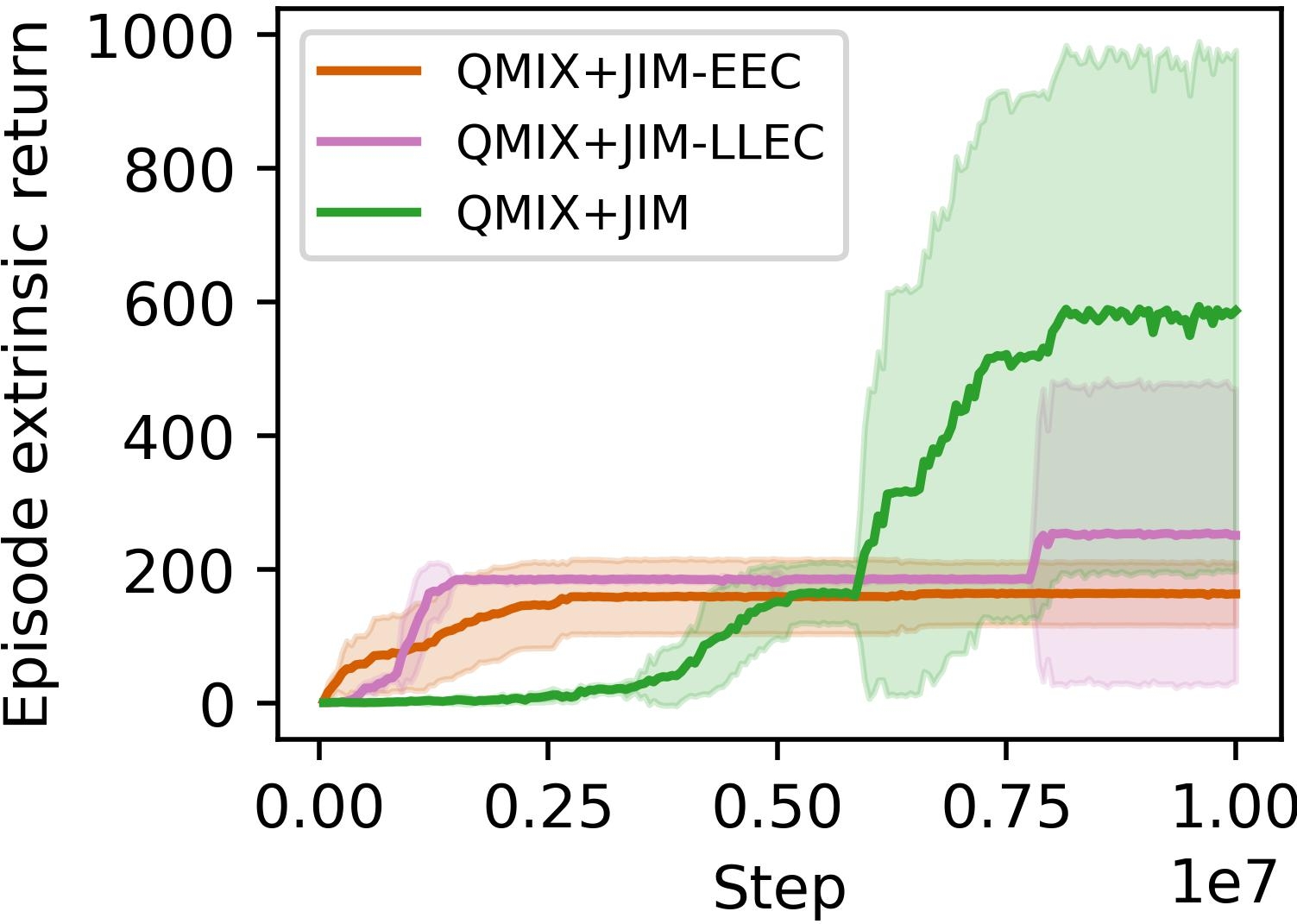}}
    \subcaptionbox{\label{fig:ablation_last}}{\includegraphics[width=0.16\textwidth]{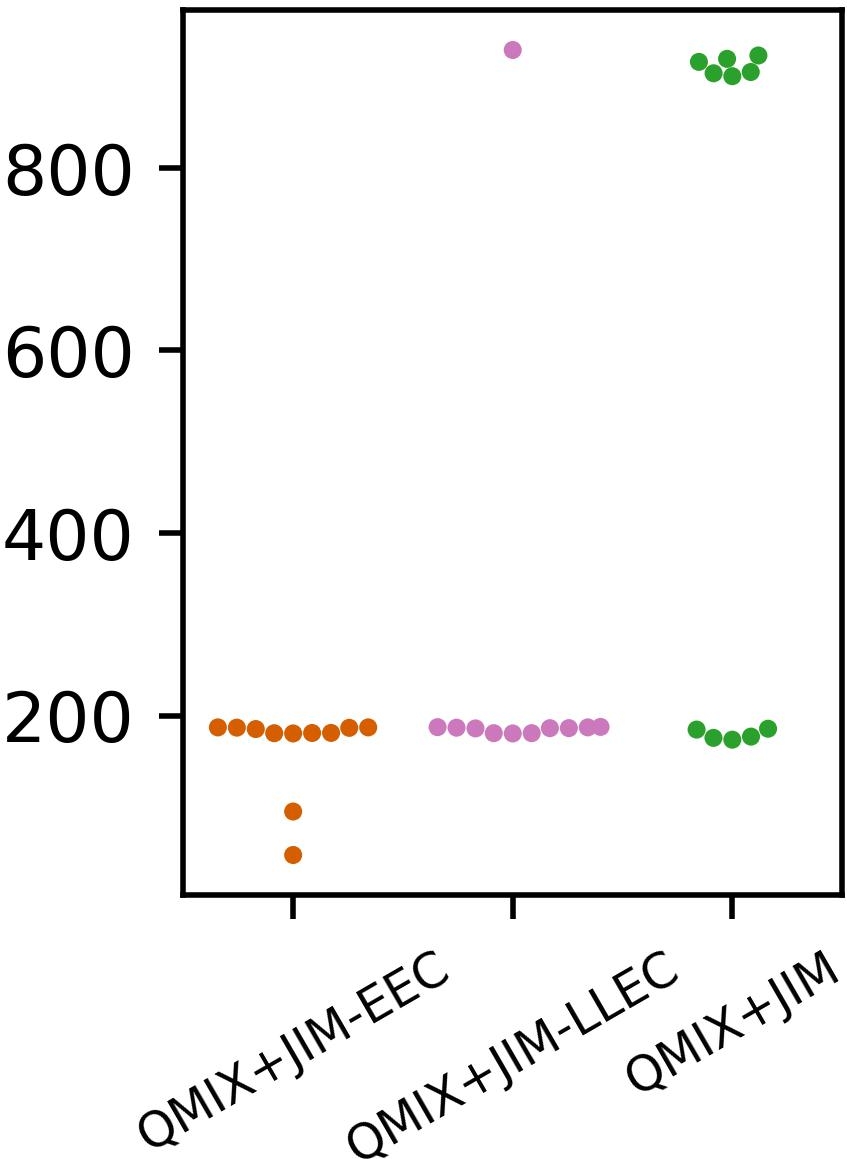}}
    \caption{Ablation study of JIM in the coordinated placement task. (a) shows the training curves with the mean and standard deviation across 11 independent runs each, while (b) displays the performance of each independent run at the last iteration of training. The two ablated versions only feature one of the two exploration criteria defined in Section \ref{sec:JIM:IntrRew}: JIM-EEC for $N_{EEC}$ and JIM-LLEC for $N_{LLEC}$. The results show the importance of combining the two criteria.}
    \Description[Ablation study in the coordinated placement task.]{Two graphs showing training curves of the ablation study in the coordinated placement task and the final reward level obtained by each independent runs (there are 11 for each trained version). Joint Intrinsic Motivation is better than its two ablated version.}
    \label{fig:ablation_results}
\end{figure}

\subsection{Ablation study}

In this ablation study, JIM is compared with two ablated versions of the reward: one with only the \textit{episodic exploration criterion} $N_{EEC}$ (JIM-EEC) and one with only the \textit{life-long exploration criterion} $N_{LLEC}$ (JIM-LLEC). Note that JIM-LLEC is actually equivalent to NovelD \cite{Zhang2021:NovelD} in this environment as the episodic restriction of NovelD (see Section \ref{sec:Background:IntrRew}) would be ineffective in a continuous environment such as MPE. To compare these three versions properly, we scale the intrinsic rewards of the two ablated models to be at a similar magnitude as the intrinsic reward generated by JIM. Figure \ref{fig:ablation_results} shows the results of training these versions in the coordinated placement task, with 11 independent runs each. Both ablated algorithms perform significantly worse than JIM. First, the episodic bonus of JIM-EEC alone lacks the motivation for discovering unseen configurations. Thus, it explores less and is not able to find the optimal solution to the task. Meanwhile, without the episodic restriction, JIM-LLEC develops less efficient exploration strategies. This confirms that, as shown in a recent study \cite{Andres2022:EvalIntrRew}, the episodic restriction implemented in NovelD and other intrinsic rewards \cite{Raileanu2020:RIDE,Badia2020:NGU} is crucial for developing efficient exploration strategies.  Overall, this proves the importance of combining the two stages of exploration defined in $N_{LLEC}$ and $N_{EEC}$. 

\subsection{Scaling up to more agents}\label{sec:Exp:scaling}

Finally, we evaluate JIM in a scenario with four agents using a modified version of the $\mathtt{rel\_overgen}$ environment introduced in Section~\ref{sec6.1} (see Appendix \ref{app:rel_overgen} for more details). As in the two-agent version, the reward function has a high reward spike in one corner of the space and a low reward plateau in the other corner, making relative overgeneralization prone to arise. With more agents, the number of dimensions of the joint observation increases linearly (in this case: from 80 dimensions  with two agents to 160 with four agents) while the number of possible states increases exponentially, making the search for the optimal strategy significantly more challenging.

\begin{figure}[t]
    \centering
    \begin{tabular}{c}
        \includegraphics[width=0.37\textwidth]{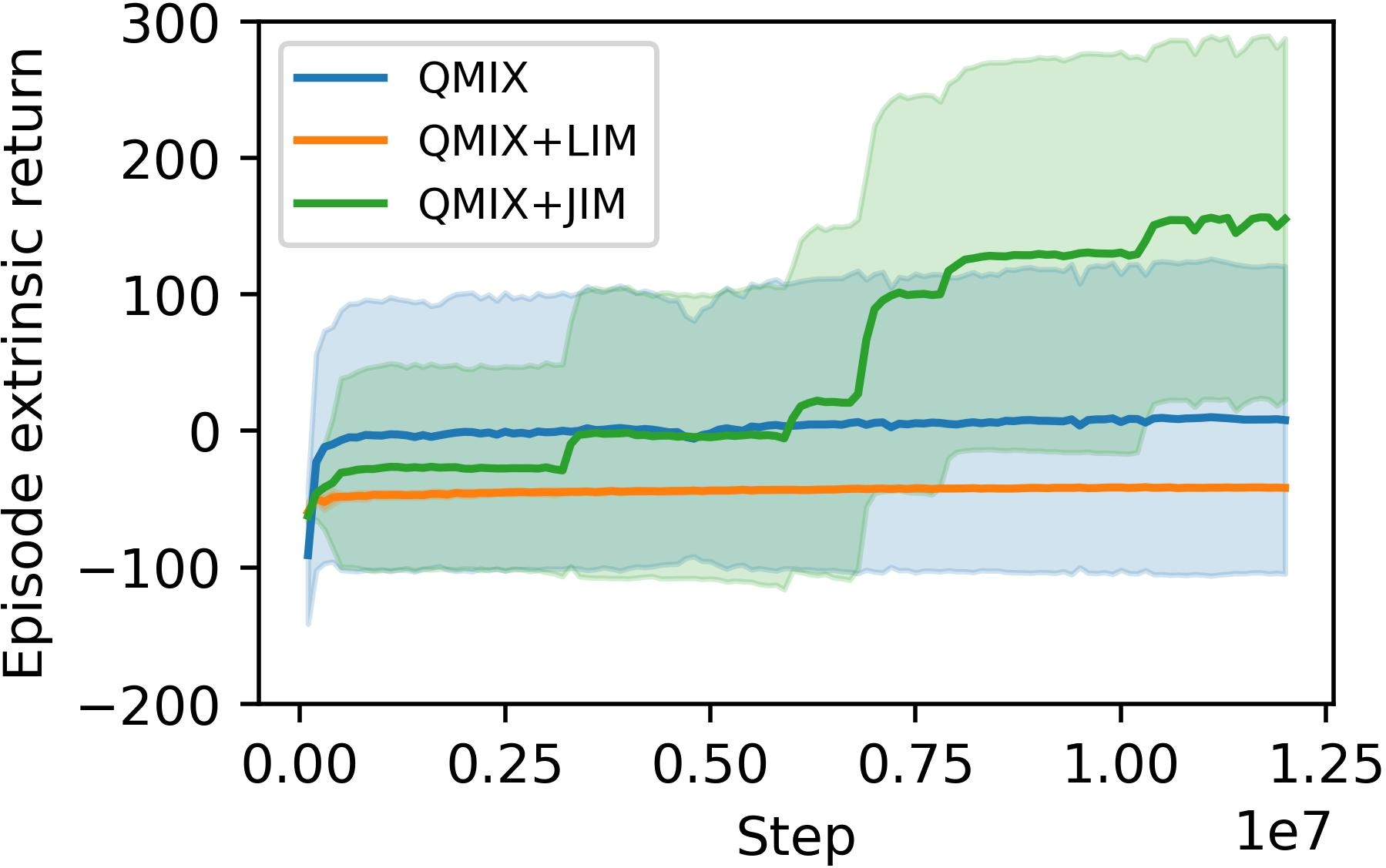}\\
        \includegraphics[width=0.37\textwidth]{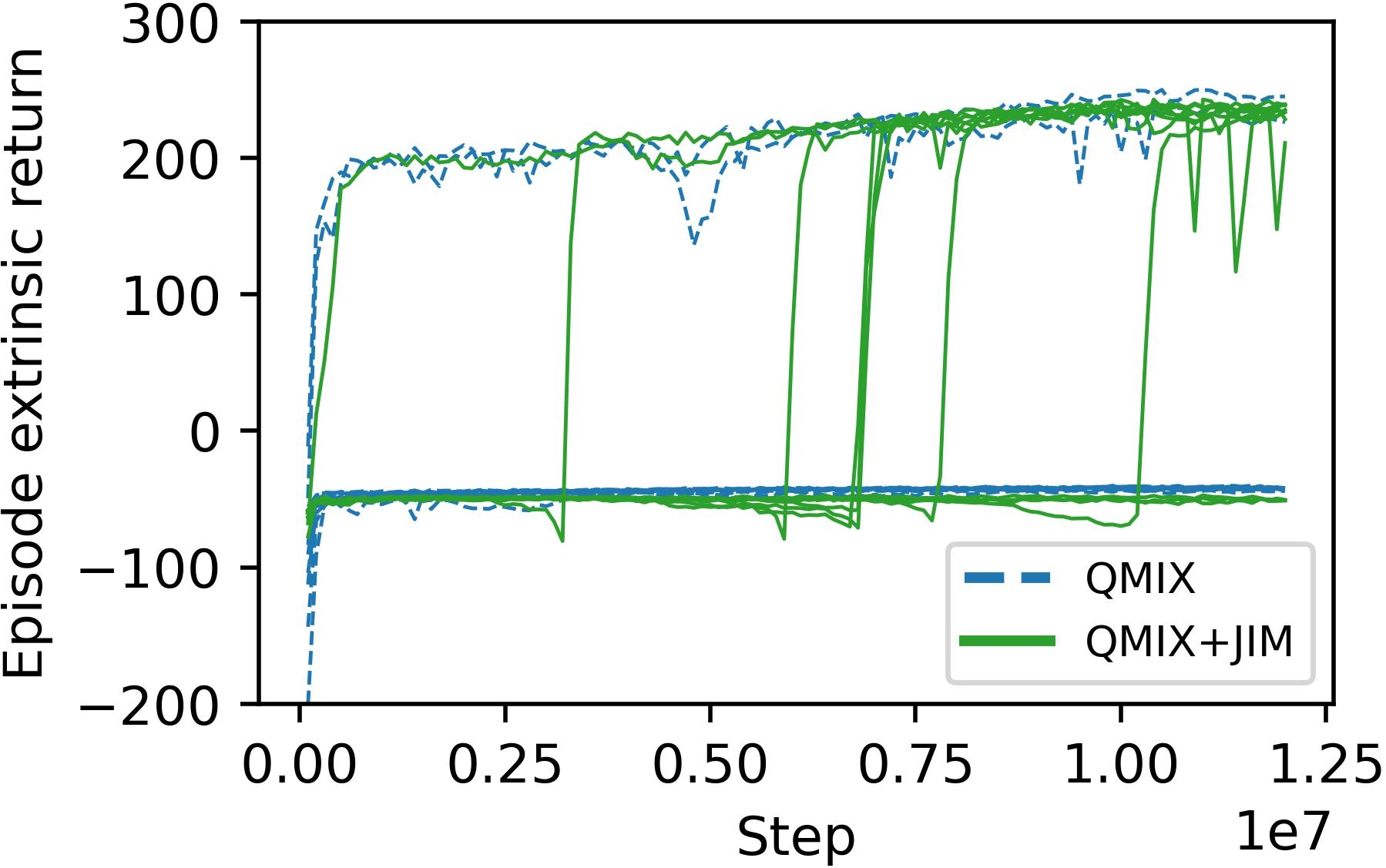}\\
    \end{tabular}
    \caption{Training performance of QMIX, QMIX+LIM and QMIX+JIM in the four-agent version of $\mathtt{rel\_overgen}$. Top graph shows the mean and standard deviation across 11 runs each, bottom graph displays all single runs (QMIX in dotted lines for clarity, QMIX+LIM is omitted as it never goes beyond the suboptimal strategy).}
    \Description[Training in the 4-agent version of the relative overgeneralization environment.]{Training in the 4-agent version of the relative overgeneralization environment. Joint Intrinsic Motivation converges to the optimal solution more often than the other two versions. }
    \label{fig:ro_4a}
\end{figure}

Figure~\ref{fig:ro_4a} shows the results in this scenario for 11 independent runs each. As with previous experiments, JIM improves the performance of QMIX by upgrading its exploration capabilities. Taking a closer look at the results reveals that QMIX and QMIX+LIM find the optimal solution in respectively $2$ and $0$ runs out of the $11$, while QMIX+JIM finds the optimal solution in $8$ out of $11$ runs in the allocated time (cf. Figure~\ref{fig:ro_4a}-bottom).  The two positive results for QMIX can be attributed to beneficial initial conditions as QMIX never reaches the optimal performance otherwise. This is not the case with QMIX+JIM, which shows robustness to initial conditions thanks to active exploration of the environment. In fact, the impact of JIM becomes evident when looking at curves from individual runs, where we consistently observe significant performance enhancements subsequent to a minor initial drop in efficiency. This phenomenon reflects a deliberate shift towards exploring novel approaches when the system would otherwise remain stagnant. This is unique to JIM, as  QMIX+LIM never succeeds in finding the optimal solution, advocating for the benefits of using a global, rather than local, intrinsic reward for exploration. 
%With more agents and a larger joint-state space, local observations become even less relevant for describing the current state of the environment.

%Without intrinsic reward, QMIX either finds the optimal strategy very fast or remains stuck in a bad strategy. With JIM, the agents find the optimal strategy more often, even late in the training process, showing that JIM induces better exploration behavior. This also shows that JIM scales well to larger joint-observation spaces. 

\section{Conclusion}

In this paper, we present an algorithm for joint intrinsic motivation (JIM), which is the first method to reward active exploration of the joint-observation space. It can be integrated to enhance any Multi-Agent Deep Reinforcement Learning algorithm that uses centralized training with decentralized execution. By combining JIM with the state-of-the-art QMIX algorithm, we demonstrate that it outperforms the original QMIX implementation as well as a modified QMIX algorithm using single-agent intrinsic rewards. We show that active exploration is a key component for multi-agent learning in environments with sparse rewards. Moreover, joint exploration enables the discovery of optimal coordinated behaviors that would be hard to find otherwise as they necessitate a high level of coordination between agents. 

This shows the importance of using joint observations in the process of computing intrinsic rewards for a multi-agent system. In fact, the joint observation is the best estimate of the global state of the environment available for the agents. Using it allows more efficient learning of multi-agent joint behaviors and is computationally less expensive than having to compute local intrinsic rewards for each agent. These results should encourage research on how joint observations can be used in other kinds of intrinsic rewards to shape the agents' behavior further. 

% \section*{Acknowledgment} TODO: use the 'acks' environment instead (see AAMAS template)

% The authors appreciate the ECE for financing the Lambda Quad Max Deep Learning server, which is employed to obtain the results illustrated in the present work. 

\balance
\bibliographystyle{ACM-Reference-Format} 
\bibliography{mypap}


\begin{thebibliography}{35}

%%% the Web address.

\ifx \showCODEN    \undefined \def \showCODEN     #1{\unskip}     \fi
\ifx \showDOI      \undefined \def \showDOI       #1{#1}\fi
\ifx \showISBNx    \undefined \def \showISBNx     #1{\unskip}     \fi
\ifx \showISBNxiii \undefined \def \showISBNxiii  #1{\unskip}     \fi
\ifx \showISSN     \undefined \def \showISSN      #1{\unskip}     \fi
\ifx \showLCCN     \undefined \def \showLCCN      #1{\unskip}     \fi
\ifx \shownote     \undefined \def \shownote      #1{#1}          \fi
\ifx \showarticletitle \undefined \def \showarticletitle #1{#1}   \fi
\ifx \showURL      \undefined \def \showURL       {\relax}        \fi
% invisible to TeX
\providecommand\bibfield[2]{#2}
\providecommand\bibinfo[2]{#2}
\providecommand\natexlab[1]{#1}
\providecommand\showeprint[2][]{arXiv:#2}

\bibitem[\protect\citeauthoryear{Andres, Villar-Rodriguez, and Ser}{Andres et~al\mbox{.}}{2022}]%
        {Andres2022:EvalIntrRew}
\bibfield{author}{\bibinfo{person}{Alain Andres}, \bibinfo{person}{Esther Villar-Rodriguez}, {and} \bibinfo{person}{Javier~Del Ser}.} \bibinfo{year}{2022}\natexlab{}.
\newblock \showarticletitle{An Evaluation Study of Intrinsic Motivation Techniques applied to Reinforcement Learning over Hard Exploration Environments}. In \bibinfo{booktitle}{\emph{arXiv:2205.11184}}.
\newblock


\bibitem[\protect\citeauthoryear{Badia, Sprechmann, Vitvitskyi, Guo, Piot, Kapturowski, Tieleman, Arjovsky, Pritzel, Bolt, and Blundell}{Badia et~al\mbox{.}}{2020}]%
        {Badia2020:NGU}
\bibfield{author}{\bibinfo{person}{Adrià~Puigdomènech Badia}, \bibinfo{person}{Pablo Sprechmann}, \bibinfo{person}{Alex Vitvitskyi}, \bibinfo{person}{Daniel Guo}, \bibinfo{person}{Bilal Piot}, \bibinfo{person}{Steven Kapturowski}, \bibinfo{person}{Olivier Tieleman}, \bibinfo{person}{Martin Arjovsky}, \bibinfo{person}{Alexander Pritzel}, \bibinfo{person}{Andrew Bolt}, {and} \bibinfo{person}{Charles Blundell}.} \bibinfo{year}{2020}\natexlab{}.
\newblock \showarticletitle{Never Give Up: Learning Directed Exploration Strategies}. In \bibinfo{booktitle}{\emph{8th International Conference on Learning Representations}}.
\newblock
\urldef\tempurl%
\url{https://openreview.net/forum?id=Sye57xStvB}
\showURL{%
\tempurl}


\bibitem[\protect\citeauthoryear{Burda, Edwards, Storkey, and Klimov}{Burda et~al\mbox{.}}{2019}]%
        {Burda2019:RND}
\bibfield{author}{\bibinfo{person}{Yuri Burda}, \bibinfo{person}{Harrison Edwards}, \bibinfo{person}{Amos Storkey}, {and} \bibinfo{person}{Oleg Klimov}.} \bibinfo{year}{2019}\natexlab{}.
\newblock \showarticletitle{Exploration by Random Network Distillation}. In \bibinfo{booktitle}{\emph{7th International Conference on Learning Representations}}.
\newblock


\bibitem[\protect\citeauthoryear{Du, Han, Fang, Dai, Liu, and Tao}{Du et~al\mbox{.}}{2019}]%
        {Du2019:LIIR}
\bibfield{author}{\bibinfo{person}{Yali Du}, \bibinfo{person}{Lei Han}, \bibinfo{person}{Meng Fang}, \bibinfo{person}{Tianhong Dai}, \bibinfo{person}{Ji Liu}, {and} \bibinfo{person}{Dacheng Tao}.} \bibinfo{year}{2019}\natexlab{}.
\newblock \showarticletitle{{LIIR}: Learning Individual Intrinsic Reward in Multi-Agent Reinforcement Learning}. In \bibinfo{booktitle}{\emph{Advances in Neural Information Processing Systems}}, Vol.~\bibinfo{volume}{32}.
\newblock
\urldef\tempurl%
\url{https://proceedings.neurips.cc/paper/2019/hash/07a9d3fed4c5ea6b17e80258dee231fa-Abstract.html}
\showURL{%
\tempurl}


\bibitem[\protect\citeauthoryear{Flet-Berliac, Ferret, Pietquin, Preux, and Geist}{Flet-Berliac et~al\mbox{.}}{2021}]%
        {FletBerliac2021:AGAC}
\bibfield{author}{\bibinfo{person}{Yannis Flet-Berliac}, \bibinfo{person}{Johan Ferret}, \bibinfo{person}{Olivier Pietquin}, \bibinfo{person}{Philippe Preux}, {and} \bibinfo{person}{Matthieu Geist}.} \bibinfo{year}{2021}\natexlab{}.
\newblock \showarticletitle{Adversarially Guided Actor-Critic}. In \bibinfo{booktitle}{\emph{9th International Conference on Learning Representations}}.
\newblock
\urldef\tempurl%
\url{https://hal.inria.fr/hal-03167169}
\showURL{%
\tempurl}


\bibitem[\protect\citeauthoryear{Foerster, Farquhar, Afouras, Nardelli, and Whiteson}{Foerster et~al\mbox{.}}{2018}]%
        {Foerster2018:COMA}
\bibfield{author}{\bibinfo{person}{Jakob Foerster}, \bibinfo{person}{Gregory Farquhar}, \bibinfo{person}{Triantafyllos Afouras}, \bibinfo{person}{Nantas Nardelli}, {and} \bibinfo{person}{Shimon Whiteson}.} \bibinfo{year}{2018}\natexlab{}.
\newblock \showarticletitle{Counterfactual Multi-Agent Policy Gradients}. In \bibinfo{booktitle}{\emph{Proceedings of the AAAI Conference on Artificial Intelligence}}, Vol.~\bibinfo{volume}{32}.
\newblock
\urldef\tempurl%
\url{https://ojs.aaai.org/index.php/AAAI/article/view/11794}
\showURL{%
\tempurl}


\bibitem[\protect\citeauthoryear{Henaff, Raileanu, Jiang, and Rockt{\"a}schel}{Henaff et~al\mbox{.}}{2022}]%
        {Henaff2022:E3B}
\bibfield{author}{\bibinfo{person}{Mikael Henaff}, \bibinfo{person}{Roberta Raileanu}, \bibinfo{person}{Minqi Jiang}, {and} \bibinfo{person}{Tim Rockt{\"a}schel}.} \bibinfo{year}{2022}\natexlab{}.
\newblock \showarticletitle{Exploration via Elliptical Episodic Bonuses}. In \bibinfo{booktitle}{\emph{Advances in Neural Information Processing Systems}}, Vol.~\bibinfo{volume}{35}. \bibinfo{pages}{37631--37646}.
\newblock
\urldef\tempurl%
\url{https://openreview.net/forum?id=Xg-yZos9qJQ}
\showURL{%
\tempurl}


\bibitem[\protect\citeauthoryear{Iqbal and Sha}{Iqbal and Sha}{2019}]%
        {Iqbal2019:MultiExplore}
\bibfield{author}{\bibinfo{person}{Shariq Iqbal} {and} \bibinfo{person}{Fei Sha}.} \bibinfo{year}{2019}\natexlab{}.
\newblock \showarticletitle{Coordinated Exploration via Intrinsic Rewards for Multi-Agent Reinforcement Learning}. In \bibinfo{booktitle}{\emph{arXiv:1905.12127}}.
\newblock
\urldef\tempurl%
\url{https://openreview.net/forum?id=rkltE0VKwH}
\showURL{%
\tempurl}


\bibitem[\protect\citeauthoryear{Jaques, Lazaridou, Hughes, Gulcehre, Ortega, Strouse, Leibo, and de~Freitas}{Jaques et~al\mbox{.}}{2019}]%
        {Jaques2019}
\bibfield{author}{\bibinfo{person}{Natasha Jaques}, \bibinfo{person}{Angeliki Lazaridou}, \bibinfo{person}{Edward Hughes}, \bibinfo{person}{Caglar Gulcehre}, \bibinfo{person}{Pedro~A. Ortega}, \bibinfo{person}{DJ Strouse}, \bibinfo{person}{Joel~Z. Leibo}, {and} \bibinfo{person}{Nando de Freitas}.} \bibinfo{year}{2019}\natexlab{}.
\newblock \showarticletitle{Social Influence as Intrinsic Motivation for Multi-Agent Deep Reinforcement Learning}. In \bibinfo{booktitle}{\emph{Proceedings of the 36th International Conference on Machine Learning}}, Vol.~\bibinfo{volume}{97}. \bibinfo{pages}{3040--3049}.
\newblock


\bibitem[\protect\citeauthoryear{Lehman and Stanley}{Lehman and Stanley}{2011}]%
        {lehman2011abandoning}
\bibfield{author}{\bibinfo{person}{Joel Lehman} {and} \bibinfo{person}{Kenneth~O. Stanley}.} \bibinfo{year}{2011}\natexlab{}.
\newblock \showarticletitle{Abandoning objectives: Evolution through the search for novelty alone}. In \bibinfo{booktitle}{\emph{Evolutionary Computation}}, Vol.~\bibinfo{volume}{19}. \bibinfo{pages}{189--223}.
\newblock


\bibitem[\protect\citeauthoryear{Lowe, Wu, Tamar, Harb, Abbeel, and Mordatch}{Lowe et~al\mbox{.}}{2017}]%
        {Lowe2017:MADDPG}
\bibfield{author}{\bibinfo{person}{Ryan Lowe}, \bibinfo{person}{Yi Wu}, \bibinfo{person}{Aviv Tamar}, \bibinfo{person}{Jean Harb}, \bibinfo{person}{Pieter Abbeel}, {and} \bibinfo{person}{Igor Mordatch}.} \bibinfo{year}{2017}\natexlab{}.
\newblock \showarticletitle{Multi-Agent Actor-Critic for Mixed Cooperative-Competitive Environments}. In \bibinfo{booktitle}{\emph{Advances in Neural Information Processing Systems}}, Vol.~\bibinfo{volume}{30}.
\newblock


\bibitem[\protect\citeauthoryear{Lupu, Cui, Hu, and Foerster}{Lupu et~al\mbox{.}}{2021}]%
        {Lupu2021:TrajeDi}
\bibfield{author}{\bibinfo{person}{Andrei Lupu}, \bibinfo{person}{Brandon Cui}, \bibinfo{person}{Hengyuan Hu}, {and} \bibinfo{person}{Jakob Foerster}.} \bibinfo{year}{2021}\natexlab{}.
\newblock \showarticletitle{Trajectory Diversity for Zero-Shot Coordination}. In \bibinfo{booktitle}{\emph{Proceedings of the 38th International Conference on Machine Learning}}, Vol.~\bibinfo{volume}{139}. \bibinfo{pages}{7204--7213}.
\newblock
\urldef\tempurl%
\url{https://proceedings.mlr.press/v139/lupu21a.html}
\showURL{%
\tempurl}


\bibitem[\protect\citeauthoryear{Ma, Wang, Fei-Fei, Bernstein, and Krishna}{Ma et~al\mbox{.}}{2022}]%
        {Ma2022:ELIGN}
\bibfield{author}{\bibinfo{person}{Zixian Ma}, \bibinfo{person}{Rose~E Wang}, \bibinfo{person}{Li Fei-Fei}, \bibinfo{person}{Michael~S. Bernstein}, {and} \bibinfo{person}{Ranjay Krishna}.} \bibinfo{year}{2022}\natexlab{}.
\newblock \showarticletitle{{ELIGN}: Expectation Alignment as a Multi-Agent Intrinsic Reward}. In \bibinfo{booktitle}{\emph{Advances in Neural Information Processing Systems}}, Vol.~\bibinfo{volume}{35}. \bibinfo{pages}{8304--8317}.
\newblock
\urldef\tempurl%
\url{https://openreview.net/forum?id=uPyNR2yPoe}
\showURL{%
\tempurl}


\bibitem[\protect\citeauthoryear{Mahajan, Rashid, Samvelyan, and Whiteson}{Mahajan et~al\mbox{.}}{2019}]%
        {Mahajan2019:MAVEN}
\bibfield{author}{\bibinfo{person}{Anuj Mahajan}, \bibinfo{person}{Tabish Rashid}, \bibinfo{person}{Mikayel Samvelyan}, {and} \bibinfo{person}{Shimon Whiteson}.} \bibinfo{year}{2019}\natexlab{}.
\newblock \showarticletitle{{MAVEN}: Multi-Agent Variational Exploration}. In \bibinfo{booktitle}{\emph{Advances in Neural Information Processing Systems}}, Vol.~\bibinfo{volume}{32}.
\newblock
\urldef\tempurl%
\url{https://proceedings.neurips.cc/paper/2019/file/f816dc0acface7498e10496222e9db10-Paper.pdf}
\showURL{%
\tempurl}


\bibitem[\protect\citeauthoryear{Mordatch and Abbeel}{Mordatch and Abbeel}{2018}]%
        {Mordatch2018}
\bibfield{author}{\bibinfo{person}{Igor Mordatch} {and} \bibinfo{person}{Pieter Abbeel}.} \bibinfo{year}{2018}\natexlab{}.
\newblock \showarticletitle{Emergence of Grounded Compositional Language in Multi-Agent Populations}. In \bibinfo{booktitle}{\emph{Proceedings of the AAAI Conference on Artificial Intelligence}}.
\newblock
\urldef\tempurl%
\url{https://www.aaai.org/ocs/index.php/AAAI/AAAI18/paper/view/17007}
\showURL{%
\tempurl}


\bibitem[\protect\citeauthoryear{Oliehoek and Amato}{Oliehoek and Amato}{2016}]%
        {Oliehoek2016:DecPOMDP}
\bibfield{author}{\bibinfo{person}{Frans~A. Oliehoek} {and} \bibinfo{person}{Christopher Amato}.} \bibinfo{year}{2016}\natexlab{}.
\newblock \bibinfo{booktitle}{\emph{A Concise Introduction to Decentralized POMDPs}}.
\newblock \bibinfo{publisher}{Springer}.
\newblock
\urldef\tempurl%
\url{https://www.ccis.northeastern.edu/home/camato/publications/OliehoekAmato16book.pdf}
\showURL{%
\tempurl}


\bibitem[\protect\citeauthoryear{OpenAI, Berner, Brockman, Chan, Cheung, Dębiak, Dennison, Farhi, Fischer, Hashme, Hesse, Józefowicz, Gray, Olsson, Pachocki, Petrov, d.~O.~Pinto, Raiman, Salimans, Schlatter, Schneider, Sidor, Sutskever, Tang, Wolski, and Zhang}{OpenAI et~al\mbox{.}}{2019}]%
        {OpenAI2019:DOTA2}
\bibfield{author}{\bibinfo{person}{OpenAI}, \bibinfo{person}{Christopher Berner}, \bibinfo{person}{Greg Brockman}, \bibinfo{person}{Brooke Chan}, \bibinfo{person}{Vicki Cheung}, \bibinfo{person}{Przemysław Dębiak}, \bibinfo{person}{Christy Dennison}, \bibinfo{person}{David Farhi}, \bibinfo{person}{Quirin Fischer}, \bibinfo{person}{Shariq Hashme}, \bibinfo{person}{Chris Hesse}, \bibinfo{person}{Rafal Józefowicz}, \bibinfo{person}{Scott Gray}, \bibinfo{person}{Catherine Olsson}, \bibinfo{person}{Jakub Pachocki}, \bibinfo{person}{Michael Petrov}, \bibinfo{person}{Henrique~P. d. O.~Pinto}, \bibinfo{person}{Jonathan Raiman}, \bibinfo{person}{Tim Salimans}, \bibinfo{person}{Jeremy Schlatter}, \bibinfo{person}{Jonas Schneider}, \bibinfo{person}{Szymon Sidor}, \bibinfo{person}{Ilya Sutskever}, \bibinfo{person}{Jie Tang}, \bibinfo{person}{Filip Wolski}, {and} \bibinfo{person}{Susan Zhang}.} \bibinfo{year}{2019}\natexlab{}.
\newblock \showarticletitle{Dota 2 with Large Scale Deep Reinforcement Learning}. In \bibinfo{booktitle}{\emph{arXiv:1912.06680}}.
\newblock


\bibitem[\protect\citeauthoryear{Oudeyer and Kaplan}{Oudeyer and Kaplan}{2007}]%
        {Oudeyer2007:IntrMotiv}
\bibfield{author}{\bibinfo{person}{Pierre-Yves Oudeyer} {and} \bibinfo{person}{Frederic Kaplan}.} \bibinfo{year}{2007}\natexlab{}.
\newblock \showarticletitle{What is Intrinsic Motivation? A Typology of Computational Approaches}. In \bibinfo{booktitle}{\emph{Frontiers in neurorobotics}}, Vol.~\bibinfo{volume}{1}. \bibinfo{pages}{6}.
\newblock
\urldef\tempurl%
\url{https://doi.org/10.3389/neuro.12.006.2007}
\showDOI{\tempurl}


\bibitem[\protect\citeauthoryear{Pathak, Agrawal, Efros, and Darrell}{Pathak et~al\mbox{.}}{2017}]%
        {Pathak2017:ICM}
\bibfield{author}{\bibinfo{person}{Deepak Pathak}, \bibinfo{person}{Pulkit Agrawal}, \bibinfo{person}{Alexei~A. Efros}, {and} \bibinfo{person}{Trevor Darrell}.} \bibinfo{year}{2017}\natexlab{}.
\newblock \showarticletitle{Curiosity-driven Exploration by Self-supervised Prediction}. In \bibinfo{booktitle}{\emph{Proceedings of the 34th International Conference on Machine Learning}}, Vol.~\bibinfo{volume}{PMLR 70}. \bibinfo{pages}{2778--2787}.
\newblock


\bibitem[\protect\citeauthoryear{Raileanu and Rocktäschel}{Raileanu and Rocktäschel}{2020}]%
        {Raileanu2020:RIDE}
\bibfield{author}{\bibinfo{person}{Roberta Raileanu} {and} \bibinfo{person}{Tim Rocktäschel}.} \bibinfo{year}{2020}\natexlab{}.
\newblock \showarticletitle{{RIDE}: Rewarding Impact-Driven Exploration for Procedurally-Generated Environments}. In \bibinfo{booktitle}{\emph{9th International Conference on Learning Representations}}.
\newblock
\urldef\tempurl%
\url{https://openreview.net/forum?id=rkg-TJBFPB}
\showURL{%
\tempurl}


\bibitem[\protect\citeauthoryear{Rashid, Farquhar, Peng, and Whiteson}{Rashid et~al\mbox{.}}{2020}]%
        {Rashid2020:WQMIX}
\bibfield{author}{\bibinfo{person}{Tabish Rashid}, \bibinfo{person}{Gregory Farquhar}, \bibinfo{person}{Bei Peng}, {and} \bibinfo{person}{Shimon Whiteson}.} \bibinfo{year}{2020}\natexlab{}.
\newblock \showarticletitle{Weighted {QMIX}: Expanding Monotonic Value Function Factorisation for Deep Multi-Agent Reinforcement Learning}. In \bibinfo{booktitle}{\emph{Advances in Neural Information Processing Systems 33}}. \bibinfo{pages}{10199--10210}.
\newblock
\urldef\tempurl%
\url{https://proceedings.neurips.cc/paper/2020/hash/73a427badebe0e32caa2e1fc7530b7f3-Abstract.html}
\showURL{%
\tempurl}


\bibitem[\protect\citeauthoryear{Rashid, Samvelyan, Schroeder, Farquhar, Foerster, and Whiteson}{Rashid et~al\mbox{.}}{2018}]%
        {Rashid2018:QMIX}
\bibfield{author}{\bibinfo{person}{Tabish Rashid}, \bibinfo{person}{Mikayel Samvelyan}, \bibinfo{person}{Christian Schroeder}, \bibinfo{person}{Gregory Farquhar}, \bibinfo{person}{Jakob Foerster}, {and} \bibinfo{person}{Shimon Whiteson}.} \bibinfo{year}{2018}\natexlab{}.
\newblock \showarticletitle{{QMIX}: Monotonic Value Function Factorisation for Deep Multi-Agent Reinforcement Learning}. In \bibinfo{booktitle}{\emph{Proceedings of the 35th International Conference on Machine Learning}}, Vol.~\bibinfo{volume}{PMLR 80}. \bibinfo{pages}{4295--4304}.
\newblock
\urldef\tempurl%
\url{http://proceedings.mlr.press/v80/rashid18a.html}
\showURL{%
\tempurl}


\bibitem[\protect\citeauthoryear{Schaul, Quan, Antonoglou, and Silver}{Schaul et~al\mbox{.}}{2016}]%
        {Schaul2016:PER}
\bibfield{author}{\bibinfo{person}{Tom Schaul}, \bibinfo{person}{John Quan}, \bibinfo{person}{Ioannis Antonoglou}, {and} \bibinfo{person}{David Silver}.} \bibinfo{year}{2016}\natexlab{}.
\newblock \showarticletitle{Prioritized Experience Replay}. In \bibinfo{booktitle}{\emph{4th International Conference on Learning Representations}}.
\newblock


\bibitem[\protect\citeauthoryear{Schmidhuber}{Schmidhuber}{1991}]%
        {Schmidhuber1991}
\bibfield{author}{\bibinfo{person}{J{\"u}rgen Schmidhuber}.} \bibinfo{year}{1991}\natexlab{}.
\newblock \showarticletitle{A Possibility for Implementing Curiosity and Boredom in Model-Building Neural Controllers}. In \bibinfo{booktitle}{\emph{Proceedings of the International Conference on Simulation of Adaptive Behavior: From Animals to Animats}}. \bibinfo{pages}{222--227}.
\newblock


\bibitem[\protect\citeauthoryear{Shalev-Shwartz, Shammah, and Shashua}{Shalev-Shwartz et~al\mbox{.}}{2016}]%
        {Shalev2016:AutonomousDriving}
\bibfield{author}{\bibinfo{person}{Shai Shalev-Shwartz}, \bibinfo{person}{Shaked Shammah}, {and} \bibinfo{person}{Amnon Shashua}.} \bibinfo{year}{2016}\natexlab{}.
\newblock \showarticletitle{Safe, Multi-Agent, Reinforcement Learning for Autonomous Driving}. In \bibinfo{booktitle}{\emph{arXiv:1610.03295}}.
\newblock


\bibitem[\protect\citeauthoryear{Son, Kim, Kang, Hostallero, and Yi}{Son et~al\mbox{.}}{2019}]%
        {Son2019:Qtran}
\bibfield{author}{\bibinfo{person}{Kyunghwan Son}, \bibinfo{person}{Daewoo Kim}, \bibinfo{person}{Wan~Ju Kang}, \bibinfo{person}{David~Earl Hostallero}, {and} \bibinfo{person}{Yung Yi}.} \bibinfo{year}{2019}\natexlab{}.
\newblock \showarticletitle{{QTRAN}: Learning to Factorize with Transformation for Cooperative Multi-Agent Reinforcement Learning}. In \bibinfo{booktitle}{\emph{Proceedings of the 36th International Conference on Machine Learning}}, Vol.~\bibinfo{volume}{PMLR 97}. \bibinfo{pages}{5887--5896}.
\newblock


\bibitem[\protect\citeauthoryear{Sunehag, Lever, Gruslys, Czarnecki, Zambaldi, Jaderberg, Lanctot, Sonnerat, Leibo, Tuyls, and Graepel}{Sunehag et~al\mbox{.}}{2018}]%
        {Sunehag2018:VDN}
\bibfield{author}{\bibinfo{person}{Peter Sunehag}, \bibinfo{person}{Guy Lever}, \bibinfo{person}{Audrunas Gruslys}, \bibinfo{person}{Wojciech~Marian Czarnecki}, \bibinfo{person}{Vinicius Zambaldi}, \bibinfo{person}{Max Jaderberg}, \bibinfo{person}{Marc Lanctot}, \bibinfo{person}{Nicolas Sonnerat}, \bibinfo{person}{Joel~Z. Leibo}, \bibinfo{person}{Karl Tuyls}, {and} \bibinfo{person}{Thore Graepel}.} \bibinfo{year}{2018}\natexlab{}.
\newblock \showarticletitle{Value-Decomposition Networks For Cooperative Multi-Agent Learning Based On Team Reward}. In \bibinfo{booktitle}{\emph{Proceedings of the 17th International Conference on Autonomous Agents and MultiAgent Systems}}. \bibinfo{pages}{2085–2087}.
\newblock


\bibitem[\protect\citeauthoryear{Tan}{Tan}{1993}]%
        {Tan1993}
\bibfield{author}{\bibinfo{person}{Ming Tan}.} \bibinfo{year}{1993}\natexlab{}.
\newblock \showarticletitle{Multi-Agent Reinforcement Learning: Independent versus Cooperative Agents}. In \bibinfo{booktitle}{\emph{Proceedings of the Tenth International Conference on Machine Learning}}. \bibinfo{pages}{330–337}.
\newblock


\bibitem[\protect\citeauthoryear{Wang, Wang, Wu, and Zhang}{Wang et~al\mbox{.}}{2020}]%
        {Wang2020:EITI}
\bibfield{author}{\bibinfo{person}{Tonghan Wang}, \bibinfo{person}{Jianhao Wang}, \bibinfo{person}{Yi Wu}, {and} \bibinfo{person}{Chongjie Zhang}.} \bibinfo{year}{2020}\natexlab{}.
\newblock \showarticletitle{Influence-Based Multi-Agent Exploration}. In \bibinfo{booktitle}{\emph{8th International Conference on Learning Representations}}.
\newblock
\urldef\tempurl%
\url{https://openreview.net/forum?id=BJgy96EYvr}
\showURL{%
\tempurl}


\bibitem[\protect\citeauthoryear{Wei and Luke}{Wei and Luke}{2016}]%
        {Wei2016:RelOvergen}
\bibfield{author}{\bibinfo{person}{Ermo Wei} {and} \bibinfo{person}{Sean Luke}.} \bibinfo{year}{2016}\natexlab{}.
\newblock \showarticletitle{Lenient Learning in Independent-Learner Stochastic Cooperative Games}. In \bibinfo{booktitle}{\emph{The Journal of Machine Learning Research}}, Vol.~\bibinfo{volume}{17}. \bibinfo{pages}{2914--2955}.
\newblock


\bibitem[\protect\citeauthoryear{Wei, Wicke, Freelan, and Luke}{Wei et~al\mbox{.}}{2018}]%
        {Wei2018:MultiSoftQ}
\bibfield{author}{\bibinfo{person}{Ermo Wei}, \bibinfo{person}{Drew Wicke}, \bibinfo{person}{David Freelan}, {and} \bibinfo{person}{Sean Luke}.} \bibinfo{year}{2018}\natexlab{}.
\newblock \showarticletitle{Multiagent Soft Q-Learning}. In \bibinfo{booktitle}{\emph{arXiv:1804.09817}}.
\newblock


\bibitem[\protect\citeauthoryear{Wiegand}{Wiegand}{2003}]%
        {Wiegand2003:RelOvergen}
\bibfield{author}{\bibinfo{person}{Rudolf~Paul Wiegand}.} \bibinfo{year}{2003}\natexlab{}.
\newblock \emph{\bibinfo{title}{An Analysis of Cooperative Coevolutionary Algorithms}}.
\newblock \bibinfo{thesistype}{Ph.D. Dissertation}. \bibinfo{school}{George Mason University}.
\newblock


\bibitem[\protect\citeauthoryear{Wooldridge}{Wooldridge}{2009}]%
        {Wooldridge2009:IntroMAS}
\bibfield{author}{\bibinfo{person}{Michael Wooldridge}.} \bibinfo{year}{2009}\natexlab{}.
\newblock \bibinfo{booktitle}{\emph{An introduction to multiagent systems}}.
\newblock \bibinfo{publisher}{John wiley \& sons}.
\newblock


\bibitem[\protect\citeauthoryear{Yu, Velu, Vinitsky, Wang, Bayen, and Wu}{Yu et~al\mbox{.}}{2021}]%
        {Yu2021:MAPPO}
\bibfield{author}{\bibinfo{person}{Chao Yu}, \bibinfo{person}{Akash Velu}, \bibinfo{person}{Eugene Vinitsky}, \bibinfo{person}{Yu Wang}, \bibinfo{person}{Alexandre Bayen}, {and} \bibinfo{person}{Yi Wu}.} \bibinfo{year}{2021}\natexlab{}.
\newblock \showarticletitle{The Surprising Effectiveness of PPO in Cooperative, Multi-Agent Games}. In \bibinfo{booktitle}{\emph{arXiv:2103.01955}}.
\newblock


\bibitem[\protect\citeauthoryear{Zhang, Xu, Wang, Wu, Keutzer, Gonzalez, and Tian}{Zhang et~al\mbox{.}}{2021}]%
        {Zhang2021:NovelD}
\bibfield{author}{\bibinfo{person}{Tianjun Zhang}, \bibinfo{person}{Huazhe Xu}, \bibinfo{person}{Xiaolong Wang}, \bibinfo{person}{Yi Wu}, \bibinfo{person}{Kurt Keutzer}, \bibinfo{person}{Joseph~E Gonzalez}, {and} \bibinfo{person}{Yuandong Tian}.} \bibinfo{year}{2021}\natexlab{}.
\newblock \showarticletitle{{NovelD}: A Simple yet Effective Exploration Criterion}. In \bibinfo{booktitle}{\emph{Advances in Neural Information Processing Systems}}, Vol.~\bibinfo{volume}{34}. \bibinfo{pages}{25217--25230}.
\newblock
\urldef\tempurl%
\url{https://proceedings.neurips.cc/paper/2021/file/d428d070622e0f4363fceae11f4a3576-Paper.pdf}
\showURL{%
\tempurl}


\end{thebibliography}

% \newpage

\onecolumn

\appendix
\addcontentsline{toc}{section}{Appendices}
\section*{Appendices}

\section{Local Intrinsic Motivation}\label{app:LIM}

\begin{figure}[h]
    \centering
    \includegraphics[width=0.6\textwidth]{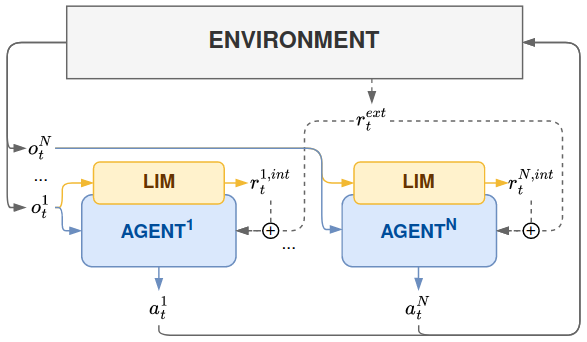}
    \caption{Architecture for local intrinsic motivation (LIM). Each agent has its own module for computing an intrinsic reward based on its local observation.}
    \Description[Diagram of the architecture of the Local Intrinsic Motivation algorithm.]{Diagram of the architecture of the Local Intrinsic Motivation algorithm. Each agent has a module for generating a local intrinsic reward based on its local observation.}
\end{figure}

\section{Hyperparameters}\label{app:hpp}

\begin{table}[h]
    \centering
    \parbox{.5\linewidth}{
        \centering
        \caption{Hyperparameters used in the $\mathtt{rel\_overgen}$ environment.}
        \begin{tabular}{|c|c|c|c|}
            \hline
            \multirow{2}{*}{Hyperparameter}               & \multicolumn{3}{c|}{Algorithm} \\ \cline{2-4} 
                                                          & JIM & JIM 4 agents & LIM \\ \hline
            Intrinsic reward weight $\beta$               & \multicolumn{3}{c|}{1}   \\ \hline
            Encoding dim $D_{\phi/\psi}$                  & 64 & 64 & 32  \\ \hline
            Hidden dim $D_{hidden}$                       & 128 & 256 & 64  \\ \hline
            Scaling factor $\alpha$                       & \multicolumn{3}{c|}{\textbf{0.5}, $0.6$}     \\ \hline
            Intrinsic reward learning rate $\alpha_{int}$ & $0.0001$ & $0.0001$, \textbf{0.0002} & $0.0001$  \\ \hline
        \end{tabular}
    }
    \hfill
    \parbox{.45\linewidth}{        
        \centering
        \caption{Hyperparameters used in the cooperative box pushing scenario.}
        \begin{tabular}{|c|cc|}
            \hline
            \multirow{2}{*}{Hyperparameter}               & \multicolumn{2}{c|}{Algorithm} \\ \cline{2-3} 
                                                          & \multicolumn{1}{c|}{JIM} & LIM \\ \hline
            Intrinsic reward weight $\beta$               & \multicolumn{1}{c|}{1}   & 1   \\ \hline
            Encoding dim $D_{\phi/\psi}$                  & \multicolumn{1}{c|}{64}  & 32  \\ \hline
            Hidden dim $D_{hidden}$                       & \multicolumn{1}{c|}{128} & 64  \\ \hline
            Scaling factor $\alpha$                       & \multicolumn{2}{c|}{$0.5$}     \\ \hline
            Intrinsic reward learning rate $\alpha_{int}$ & \multicolumn{2}{c|}{$0.0001$}  \\ \hline
        \end{tabular}
    }    
\end{table}

\begin{table}[h]
    \centering
    \caption{Hyperparameters used in the coordinated placement scenario.}
    \begin{tabular}{|c|cccc|}
        \hline
        \multirow{2}{*}{Hyperparameter}               & \multicolumn{4}{c|}{Algorithm}                                                                        \\ \cline{2-5} 
                                                      & \multicolumn{1}{c|}{JIM}     & \multicolumn{1}{c|}{LIM}     & \multicolumn{1}{c|}{JIM-LLEC} & JIM-EEC \\ \hline
        Intrinsic reward weight $\beta$               & \multicolumn{1}{c|}{1, \textbf{2}, 4} & \multicolumn{1}{c|}{1, \textbf{4}, 8} & \multicolumn{1}{c|}{1, \textbf{3}}     & \textbf{0.1}, 1  \\ \hline
        Encoding dim $D_{\phi/\psi}$                  & \multicolumn{1}{c|}{64}      & \multicolumn{1}{c|}{36}      & \multicolumn{1}{c|}{64}       & 64      \\ \hline
        Hidden dim $D_{hidden}$                       & \multicolumn{1}{c|}{512}     & \multicolumn{1}{c|}{256}     & \multicolumn{1}{c|}{512}      & 512     \\ \hline
        Scaling factor $\alpha$                       & \multicolumn{4}{c|}{$0.5$}                                                                            \\ \hline
        Intrinsic reward learning rate $\alpha_{int}$ & \multicolumn{4}{c|}{$0.0001$}                                                                         \\ \hline
    \end{tabular}
\end{table}

Tables 1 to 3 present the values chosen for hyperparameters. We only list hyperparameters specific to the intrinsic reward module, as for QMIX we use the default hyperparameters described in the original paper [22]. When we performed some search over a specific hyperparameter, we put the list of all the values we tried and put the best one in bold. Here, we detail the different parameters presented:
\begin{itemize}
    \item Intrinsic reward weight $\beta$: weight of the intrinsic reward $r_i$ against the extrinsic reward $r_e$ in the reward given to agents: $r_t=r^e_t+\beta r^{int}_t$. 
    \item Encoding dimension $D_{\phi/\psi}$: dimension of the output of the embedding networks $\phi$ and $\psi$ used in $N_{LLEC}$ and $N_{EEC}$ respectively (see Section 3).
    \item Hidden dimension $D_{hidden}$: dimension of the hidden layers in the embedding network $\phi$ and $\psi$ used in $N_{LLEC}$ and $N_{EEC}$ respectively.
    \item Scaling factor $\alpha$: parameter used in the definition of $N_{LLEC}$ (see Eq. (6)). It controls how much novelty gain we want agents to find between each step.
    \item Intrinsic reward learning rate $\alpha_{int}$: learning rate used for training the intrinsic reward module.
\end{itemize}

\section{Details on custom tasks in the Multi-agent Particle Environment}\label{app:mpe}

\subsection{Cooperative box pushing}

\begin{figure}[h]
    \centering
    \setlength{\fboxsep}{0pt}
    \setlength{\fboxrule}{1pt}
    \fbox{\includegraphics[width=0.4\textwidth]{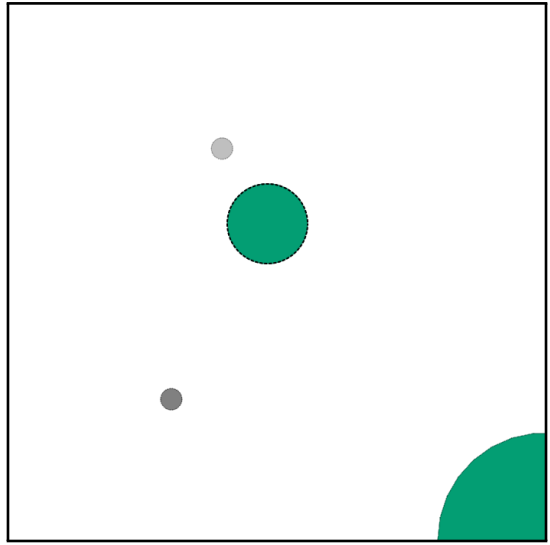}}
    \caption{Cooperative box pushing task, agents are the small grey circles, the green circle in the middle is an object to deliver to the landmark in the bottom right corner.}
    \Description[Cooperative box pushing task in the Multi-agent Particle Environment.]{Screenshots of the cooperative box pushing task, where two agents navigate around an object that can be pushed toward a landmark.}
\end{figure}

The cooperative box pushing task requires pushing a round object and placing it on top of a landmark. The environment is a 2x2 meter area with walls on the sides that block entities from moving away. At each time step, an agent gets as observation:
\begin{itemize}
    \item its personal data: its position and velocity $\text{pos}_{\text{self},x},\text{pos}_{\text{self},y},\text{vel}_{\text{self},x},\text{vel}_{\text{self},y}$,
    \item data about the other agent: a boolean indicating if the other agent is visible or not, the relative position, and the velocity of this agent $\text{is\_visible}_{\text{agent}},\text{dist}_{\text{agent},x},\text{dist}_{\text{agent},y},\text{vel}_{\text{agent},x},\text{vel}_{\text{agent},y}$,
    \item data about the object: a boolean indicating if the object is visible or not, the relative position, and the velocity of this object: $\text{is\_visible}_{\text{object}},\text{dist}_{\text{object},x},\text{dist}_{\text{object},y},\text{vel}_{\text{object},x},\text{vel}_{\text{object},y}$,
    \item and data about the landmark: a boolean indicating if the landmark is visible or not and the number of the corner it is located into (from 1 to 4): $\text{is\_visible}_{\text{landmark}},\text{corner}_{\text{landmark}}$.
\end{itemize}
Thus, the observation is a vector of dimension 16 containing this information. Relative positions of other entities (agent or object) are actually the distance to the agent, normalized by their range of observation, i.e., 
$$\text{dist}_{\text{agent},x}=\frac{\text{pos}_{\text{agent},x}-\text{pos}_{\text{self},x}}{\text{obs\_range}}.$$
The environment can be either fully observable or partially observable. In the latter case, agents have a range of observation of 60 centimeters around them. When agents or objects are outside this range, their relative position is masked with ones and their velocity with zeros. When the landmark is outside the range of observation, the corner number is masked with zero. In the fully observable case, the observation range is set to 2.83 meters, i.e., the largest distance possible to have in this 2x2 meter area.

The reward function is very sparse. At each time step, agents receive a penalty of 0.1, plus a penalty of 2 if there is a collision between agents. If the task is completed, i.e., the center of the object is placed in the area of the landmark, the agents get a reward of 100 and the episode ends.

At the start of each episode, the positions of all entities in the environment are randomly set: the agents and the object are randomly placed inside the environment, and the landmark is placed in one of the four corners of the map.

\clearpage
\subsection{Coordinated placement}

\begin{figure}[h]
    \centering
    \setlength{\fboxsep}{0pt}
    \setlength{\fboxrule}{1pt}
    \fbox{\includegraphics[width=0.4\textwidth]{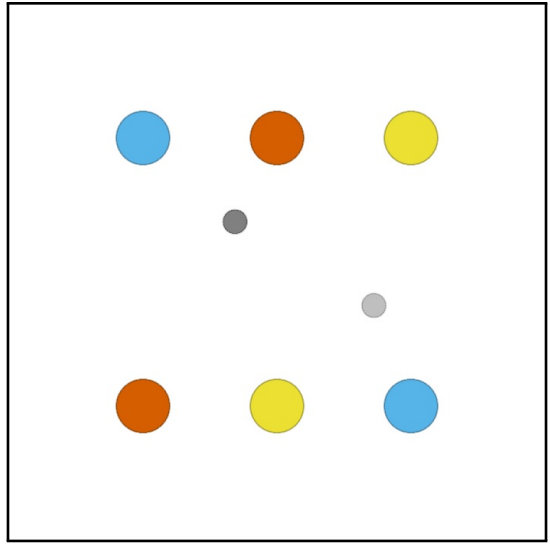}}
    \caption{Coordinated placement task, agents are the small grey circles, the colored circles represent landmarks the agents have to navigate on to gain rewards.}
    \Description[Coordinated placement tasks in the Multi-agent Particle Environment.]{Screenshots of the coordinated placement task, with two agents and two sets of three colored landmarks.}
    \label{fig:coord_place}
\end{figure}

The coordinated placement task requires navigating on top of landmarks and choosing the right landmark colors in order to maximize the obtained reward. The environment is a 2x2 meter area with walls on the sides that block entities from moving away. At each time step, an agent gets as observation:
\begin{itemize}
    \item its personal data: its position and velocity $\text{pos}_{\text{self},x},\text{pos}_{\text{self},y},\text{vel}_{\text{self},x},\text{vel}_{\text{self},y}$,
    \item data about the other agent: a boolean indicating if the other agent is visible or not, the relative position, and the velocity of this agent $\text{is\_visible}_{\text{agent}},\text{dist}_{\text{agent},x},\text{dist}_{\text{agent},y},\text{vel}_{\text{agent},x},\text{vel}_{\text{agent},y},$,
    \item for each landmark in the environment: a boolean indicating if the landmark is visible or not, the relative position of this landmark, and its color as a one-hot encoding: $\text{is\_visible}_{\text{landmark}},\text{dist}_{\text{landmark},x},\text{dist}_{\text{landmark},y},\text{is\_red},\text{is\_blue},\text{is\_yellow}$.
\end{itemize}
Thus, the observation is a vector of dimension 43 containing this information. Relative positions of other entities (agent or landmarks) are actually the distance to the agent, normalized by their range of observation, i.e., 
$$\text{dist}_{\text{agent},x}=\frac{\text{pos}_{\text{agent},x}-\text{pos}_{\text{self},x}}{\text{obs\_range}}.$$
This scenario is partially observable. Agents have a range of observation of 60 centimeters around them. When agents or objects are outside this range, their relative position is masked with ones and their velocity with zeros. For landmarks outside of the observation range, the color is masked with zeros.

The reward given at each time step depends only on which landmark has an agent placed on top:
\begin{itemize}
    \item if two agents are placed on top of red landmarks, $r^e_t=10$,
    \item if two agents are placed on top of blue landmarks, $r^e_t=2$,
    \item if two agents are placed on top of yellow landmarks, $r^e_t=1$,
    \item if only one agent is placed on top of either a blue or yellow landmark, $r^e_t=0.5$,
    \item else, $r^e_t=0$.
\end{itemize}

At the start of each episode, agents are placed randomly on the horizontal line in the middle of the map, i.e., $\text{pos}=(x=\text{uniform}(-1,1),y=0)$. The landmarks are always in the same positions, with the colors in the same order, as displayed in Figure \ref{fig:coord_place}.

\clearpage
\section{Execution times}\label{app:train_time}

\begin{figure}[h]
    \centering
    \includegraphics[width=0.7\textwidth]{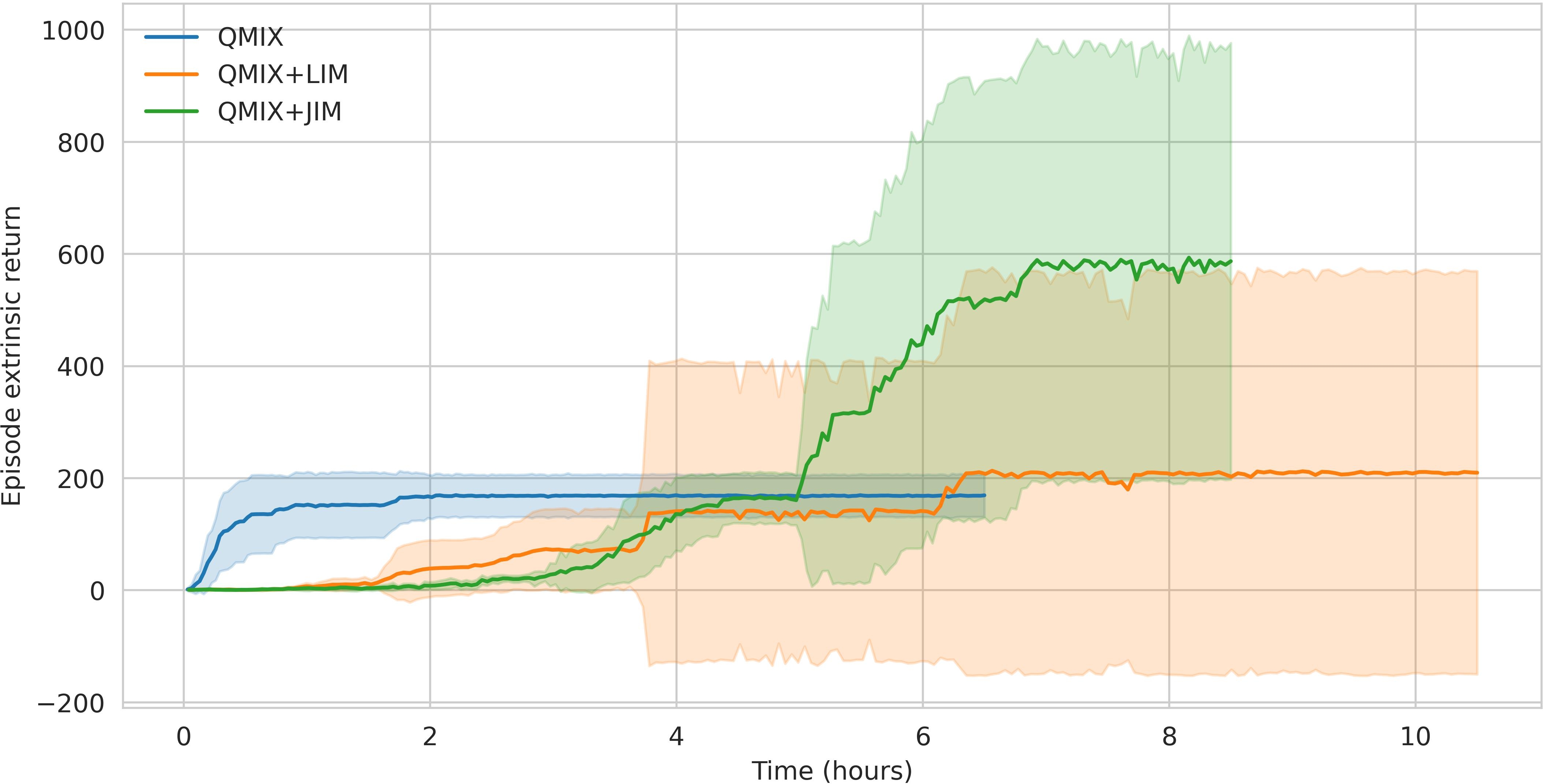}
    \caption{Training curves of QMIX, QMIX+LIM, and QMIX+JIM in the coordinated placement task with execution time on the x-axis. QMIX takes on average $6.5$ hours to train during 10 million steps, while QMIX+JIM takes $8.5$ hours and QMIX+LIM $10.5$ hours. }
    \Description{Comparison of the time taken to train QMIX, QMIX+LIM and QMIX+JIM.}
\end{figure}

\section{N-agent $\mathtt{rel\_overgen}$ environment}\label{app:rel_overgen}

To study the problem of relative overgeneralization with more than two agents, we extend the $\mathtt{rel\_overgen}$ environment to accept $N$ agents. To do so, we modify the reward definition given in the paper (see Eq. (9)): 
$$
r^{\text{ext}}_t(\mathbf{p};\delta)=\mathrm{max}\left(R^+-\frac{\delta}{D}\sum_{i=0}^{N}(p_i-r^+_i)^2,R^--\frac{1}{8D}\sum_{i=0}^{N}(p_i-r^-_i)^2\right),
$$
with $\mathbf{p}=\{p_i\}_{0<i\leq N}$ the positions of the agents, $\delta$ the coefficient controlling the size of the optimal reward spike, $D$ the dimension of each agent's state, $R^+$ the maximum value of the optimal reward spike placed at position $\mathbf{r}^+=\{r^+_i\}_{0<i\leq N}$ and $R^-$ the maximum value of the suboptimal plateau placed at position $\mathbf{r}^-=\{r^-_i\}_{0<i\leq N}$. This formula yields the same results as the two-dimensional examples shown in the paper but in a N-dimensional space.

Adding agents increases the complexity of the task exponentially. To compensate for this, we have to make the optimal reward spike larger for the task to be solvable by QMIX. In the 4-agent experiments, we use $\delta=0.9$. 

Finally, with four agents we had to lower the initial value of the $\epsilon$ parameter of QMIX for its $\epsilon$-greedy strategy. We found that increasing the number of agents led to bad results with the default $0.3$ initial value of $\epsilon$. With this hyperparameter set to $0.1$, the results were significantly better. This is likely due to the fact that agents choose separately if they explore or exploit (at least that is the case in our implementation), meaning that increasing the number of agents leads to having more randomness in the selection of each joint action. 

\end{document}